\documentclass{aa}  
\usepackage{graphicx}
\usepackage[varg]{txfonts}
\usepackage[colorlinks = true,
            linkcolor = blue,
            urlcolor  = blue,
            citecolor = blue,
            anchorcolor = blue]{hyperref}
\newcommand{\eq}[1]{Equation (\ref{eq:#1})}
\newcommand{\eqs}[2]{Equations (\ref{eq:#1}) and (\ref{eq:#2})}
\newcommand{\Eq}[1]{Equation (\ref{eq:#1})}
\newcommand{\fg}[1]{Fig.\ \ref{fig:#1}}
\newcommand{\fgs}[2]{Figs.\ \ref{fig:#1} and \ref{fig:#2}}

\newcommand{\Tb}[1]{Table\ \ref{tab:#1}}

\newcommand{\se}[1]{Section\ \ref{sec:#1}}

\newcommand{\noeq}[2]{$#1$$#2$}

\newcommand{\?}{\textcolor{red}{$^\textbf{??}$}}
\renewcommand{\?}{}

\newcommand{\comm}[1]{\textcolor{red}{[#1]}}

\newcommand{\changed}[1]{\textbf{#1}}
\renewcommand{\changed}[1]{#1}

\renewcommand{\comm}[1]{}

\begin{document} 

   \title{On the growth of pebble-accreting planetesimals}

   \author{Rico G. Visser
          \and
          Chris W. Ormel
          }

   \institute{Anton Pannekoek institute for Astronomy (API), University of Amsterdam,
              Science Park 904, 1090GE Amsterdam\\
              \email{r.g.visser@uva.nl, c.w.ormel@uva.nl}
             }

   \date{\today}

% \abstract{}{}{}{}{} 
% 5 {} token are mandatory
 
  \abstract
   {Pebble accretion is a new mechanism to quickly grow the cores of planets. In pebble accretion, gravity and gas drag conspire to yield large collisional cross sections for small particles in protoplanetary disks.
   % where a planetesimal sweeps up dust particles (pebbles) in the protoplanetary disk. This mechanism shows that under certain conditions,  gravitational but gas drag mediated interactions enable rapid growth of planetesimals to planets. 
   However, before pebble accretion commences, aerodynamical deflection may act to prevent planetesimals from becoming large, because particles tend to follow gas streamlines.}
   {We derive the planetesimal radius where pebble accretion is initiated and determine the growth timescales of planetesimals by sweepup of small particles.% fast planetesimals can grow,  the lifetime of the protoplanetary disk into consideration.
   }
   {The equation of motion for a pebble, including gas drag and gravitational interactions, is integrated in three dimensions at distances 1, 3 and 10 AU from the star.  We obtain the collision efficiency factor as the ratio of the numerically-obtained collisional cross section to the planetesimal surface area, from which we obtain the growth timescales. Integrations are conducted in the potential flow limit  (steady, inviscid) and in the Stokes flow regime (steady, viscid).}
   {Only particles of stopping time $t_s \ll t_X$ where $t_X\approx10^3$ s experience aerodynamic deflection. Even in that case, the planetesimal's gravity always ensures positive collision factors. The maximum growth timescale occurs typically at around $R\approx100 \ \mathrm{km}$, but is less for colder disks, corresponding to interactions shifting to the Safronov focusing regime. For particles $t_s \gg t_X$ pebble accretion commences only after this phase and is characterized by a steep drop in growth timescales.  %Thus, we have obtained a continuous transition between the geometrical regime and the onset of pebble accretion.  
   At 1 AU, growth timescales lie below the disk lifetime for pebbles larger than $0.03\ \mathrm{cm}$. With increasing disk orbital radius (ballistic) growth timescales increase, as well as the planetesimal radius $R_\mathrm{PA}$ where pebble accretion commences.
   %The maximum growth timescale thus increases with disk orbital radius, but this increase is mitigated by particle settling and a higher disk mass. 
   \changed{Consequently, at distances beyond \noeq{\sim}{10} AU sweepup growth timescales are always longer than $10$ Myr, while in the inner disk (\noeq{\lesssim}{3} AU) the viability of the sweepup scenario is determined by the outcome of pebble-planetesimal collisions in the geometric regime.}
   %On the other hand, effects as particle settling and an increased disk mass will decrease growth timescale. For 3 AU these effects help to bring growth timescales below $10$ Myr, but beyond \noeq{\sim}{10} AU  sweepup growth will probably exceed the lifetime of the disk. 
   %at 3 and 10 AU orbital distance,. However, a slight reduction of headwind velocity and an increase in gas and pebble densities lower peak growth times, leading to a more optimistic growth scenario. 
   We present analytical fits for the collision efficiency factors and the minimum planetesimal size needed for pebble accretion.}
   {}


   \maketitle
%
%________________________________________________________________

\section{Introduction}
\label{sec:intro}
Pebble accretion (PA) is an accretion mechanism involving both gravitational and dissipative (gas drag) forces. Thus, PA covers interactions between a big body (planetesimal or planet) and small particles in the presence of gas. During their accretion, pebbles settle in the gravitational field of the big body, just like sand grains sediment to the bottom of the ocean.\footnote{For this reason, \citet{OrmelKlahr2010} referred to these interactions as `settling' and the corresponding parameter space where they occur as the `settling regime'. We use this phrasing as a synonym with `pebble accretion', coined by  \citet{LambrechtsJohansen2012}.}
This contrasts gas-free encounters, which rely on hitting the surface of the big body. As first demonstrated by \citet{OrmelKlahr2010}, PA rates can be much higher than the corresponding gas-free limit, since cross section can become as large as the Hill sphere \citep{LambrechtsJohansen2012}. Provided the existence of a massive pebble reservoir, this virtually eradicates the timescale problem that (giant) planet formation faces in the classical scenario of planetesimal accretion.  What makes PA especially attractive is that  protoplanetary disks are inferred to harbour large amounts of pebble-size particles, as suggested by radio observations \citep{Andrews2009,Ricci2010,Testi2014}. Furthermore, a strong concentration of these particles will trigger the streaming instability, which leads to gravitationally bound objects that can produce planetesimals \citep{YoudinGoodman2005,JohansenEtal2009} -- the `seeds' required for the PA mechanism.

Recent works have advanced PA theory and applied the concept towards planet formation in the solar system and exoplanetary systems \citep{BitschEtal2015}. Compared to planetesimals, pebbles are quite mobile and drift towards the central star \citep{Weidenschilling1977}. Therefore, PA is not an isolated problem, but requires global solutions for the pebble density in the disk as function of time \citep{LambrechtsJohansen2014,BirnstielEtal2012i,KrijtEtal2015i}. How PA proceeds also depends on the size distribution of big bodies (planetesimal, protoplanets); and to reproduce the solar system architecture it is essential that only the biggest bodies in the distribution `profit' from PA   \citep{KretkeLevison2014,Levison2015}. PA ceases when planets become massive enough to alter the gas structure of the protoplanetary disk  -- the pebble isolation mass --  which, applied to the solar system, is consistent with the heavy element contents of the giant planets \citep{LambrechtsEtal2014}. In the inner solar system, PA has been invoked to explain the small size of Mars-sizes \citep{MorbidelliEtal2015} and it has been applied to explain the asteroid size distribution \citep{JohansenEtal2015}.

%argued that `chondrule accretion' explains the size distribution of asteroidal bodies. 

%However,   also attempted to reproduce the outer solar system architecture, starting from smaller (planetesimal) bodies. This proved to be a challenge, as too many intermediate-mass bodies were produced, but the same authors recently reported a breakthrough. Finally, \citet{BitschEtal2015} highlighted the advantages of PA in an exoplanet population synthesis model.

%introduced the concept of the pebble isolation mass, which is the point where . %Given suitable initial conditions\? this\? could reproduced the architecture of the gas giants in the solar system.

Under certain conditions PA results in high accretion rates. But that does not necessarily imply that PA is efficient, as particles may drift past the (proto)planet before they can be accreted \citep{OrmelKobayashi2012,Chambers2014,GuillotEtal2014,SatoEtal2015}. In particular, \mbox{\citet{GuillotEtal2014}} calculated the probability of a collision for a particle traversing  a planetesimals belt. This resulted in complex dependencies on the model parameters, but, generally, this filtering by planetesimals was seen to be most efficient for large planetesimals (protoplanets, operating in the settling regime) or for small planetesimals (which profit from a high surface area-to-mass ratio) and for weak turbulence levels (which confines particles to the midplane). On the other hand, intermediate-size planetesimals collect solids poorly. In addition, \citet{GuillotEtal2014} highlighted the importance of hydrodynamic flow effects, where small particles are so tightly coupled that they follow the streamlines of the gas, avoiding accretion by planetesimals \citep{MichaelNorey1969,Whipple1972,Slinn1976,SekiyaTakeda2003,SellentinEtal2013}. Physically, aerodynamic deflection occurs when the particle stopping time $t_s$ (which equals the particles momentum divided by the gas drag force) becomes less than the time it takes to cross the big body; that is, when the Stokes number\footnote{Remark that the Stokes number defined here differs from what is usually adopted in the astrophysical literature, where it is instead identified with the $\tau_s=t_s\Omega$ where $\Omega$ is the local orbital frequency.}
\begin{equation}
    \mathrm{St} = \frac{t_s v_\mathrm{hw}}{R}
\label{eq:Stnr}
\end{equation}
becomes less than unity, where $R$ is the planetesimal radius and $v_\mathrm{hw}$ the `headwind' velocity of the gas, which the planetesimal faces and with which the tightly coupled particles approach it.

\citet{GuillotEtal2014} conjectured that this aerodynamic deflection suppresses accretion of small particles by planetesimals. However, planetesimals do exert a gravitational force on the pebbles, which may mitigate the effect. By numerically integrating the pebble equation of motion, it is the goal of this work to quantify how severe the suppression becomes. % and to the point where gravitational forces will start to dominate over hydrodynamic drag forces for planetesimals. 
To this extent, we will calculate the mass-doubling timescales for planetesimals embedded in a sea of pebbles, identify the planetesimal size where growth experience its strongest bottleneck, and the size where PA commences (these two sizes do not necessarily coincide).

 We carry out integrations for both Stokes (high viscosity or zero Reynolds number) flow and inviscid (potential) flow. A caveat is that only steady flow patterns are considered, whereas planetesimals in the protoplanetary disk typically have Reynolds number $\gg$1, implying that the flow structure is turbulent. However, by carrying out integrations for both the Re $=0$ and the inviscid limits, we anticipate to bracket the uncertainty in the results to some extent.

We will show that, by virtue of gravitational settling, planetesimal accretion rates are always larger than zero, even for the smallest grains (although in that case the collisional cross section will be much less than the geometrical cross section). These results agree with the Stokes flow integrations by \citet{JohansenEtal2015}. Physically-motivated expressions are presented that fit our numerical results, which are valid as long as Keplerian shear effects are unimportant (i.e., small planetesimals or small particles). In particular, we derive a critical stopping time of $\approx10^3$ s, above which particles avoid aerodynamic deflection.

The plan of the paper is as follows. In \se{Model} we discuss the model setup and assumptions, including the adopted disk model and the dynamics of  gas and pebbles, followed by the numerical methods. In \se{Results} we present the collisional efficiency factors and growth timescales as function of planetesimal size, pebble size, and location in the disk. In \se{analysis}, analytical fits to the numerical data are motivated and discussed. In \se{Discussion} we discuss some implications of our results and we list our key conclusions in \se{Conclusions}.% we list our main conclusions. 

\section{Model setup and assumptions}
\label{sec:Model}
\subsection{Disk model}
The temperature and density profiles for the protoplanetary disk follow from the Minimum Mass Solar Nebula (MMSN; \citealt{Weidenschilling1977B,HayashiEtal1985}): 
\begin{eqnarray}
T(r) &=& 300\ \mathrm{K}\left ( \frac{r}{1 \ \mathrm{AU}} \right )^{-1/2},
\label{eq:Temp}\\
%\end{equation}
%\begin{equation}
\Sigma(r) &=& 1\,700\ \mathrm{g\ cm^{-2}}\left ( \frac{r}{1 \ \mathrm{AU}} \right )^{-3/2},
\label{eq:surfdens}
\end{eqnarray}
with $r$ the distance from the star. If we assume hydrostatic equilibrium and an isothermal temperature profile in the vertical direction, the density profile is a Gaussian:\begin{equation}
\rho(r,z) =\frac{\Sigma (r) }{H\sqrt{2\pi}} \exp\left[{-\frac{1}{2}\left ( \frac{z}{H} \right )^2}\right],
\label{eq:density}
\end{equation}
with $z$ the vertical distance and the scaleheight $H = c_{\mathrm{s}}/\Omega_0$, $\Omega_0$ the local Keplerian frequency and  $c_{s} =\sqrt{k_{b}T/\bar{m}}$, the isothermal sound speed, $k_b$ Boltzmann's constant and $\bar{m} = 2.34 \times 10^{-24} \ \mathrm{g}$ the mean molecular weight. Since the gas in the disk is partially pressure supported, it rotates at speeds slightly below the Kepler speed. The planetesimal thus faces a headwind of magnitude \citep{Nakagawa1986}:
\begin{equation}
    v_\mathrm{hw} = \frac{c_s^2}{2v_k} \frac{\partial \log P}{\partial \log r}= 5780  \ \mathrm{cm \ s^{-1}},
    \label{eq:headwind}
\end{equation} 
with $P(r)$ the pressure of the gas. For power-law profiles of $T$ and $\Sigma$ the headwind velocity is independent of disk orbital radius $r$. \changed{Note that the MMSN profile has been adopted for convenience; more realistic disk models give rise to different profiles than \eqs{Temp}{surfdens}. For example, the hydrodynamical simulations of \citet{BitschEtal2015}, which include radiative transport, result in a headwind velocity that is lower than \eq{headwind}.  In \se{disk-model}, we briefly consider alternative disk profiles. We note, however, that the dimensionless fit formulas that we present in \se{analysis} cover every disk profile.}

%; $v_\mathrm{hw}$ only depends on $
%therefore the numerical value in \Eq{headwind} is valid for all simulations.

\subsection{The pebble equation of motion}
For the integration of pebble streamlines we choose a local frame, co-moving with the planetesimal. The planetesimal is assumed to be massive and to move on a circular Keplerian orbit. The equation of motion in three dimensions is given by \citep{OrmelKlahr2010}:
\begin{equation}
    \frac{\mathrm{d}\mathbf{v}}{\mathrm{d}t} = 
    \begin{pmatrix}
         2\Omega_0 v_y + 3\Omega_0^2x\\ 
        -2\Omega_0v_x\\ 
         0
    \end{pmatrix} 
    -\frac{GM}{r^{3}}
    \begin{pmatrix}
        x\\ 
        y\\ 
        z
    \end{pmatrix} 
    + \mathbf{F}_\mathrm{drag},
\label{eq:eqofmotion}
\end{equation} 
with $M$ the mass of the planetesimal $r$ the distance from the local origin and $x , y, z$ the coordinates of the pebble in the local frame. The first terms of \eq{eqofmotion} consist of the Coriolis acceleration and tidal accelerations respectively, while the second term describes the planetesimal gravity. Note that the stellar acceleration term in the $z$-direction $(-\Omega_0^2 z)$, has been omitted as this would otherwise render vertical impact parameters infinite for a laminar gas flow adopted here.

\begin{figure}[t]
    \centering
    \includegraphics[width=0.47\textwidth]{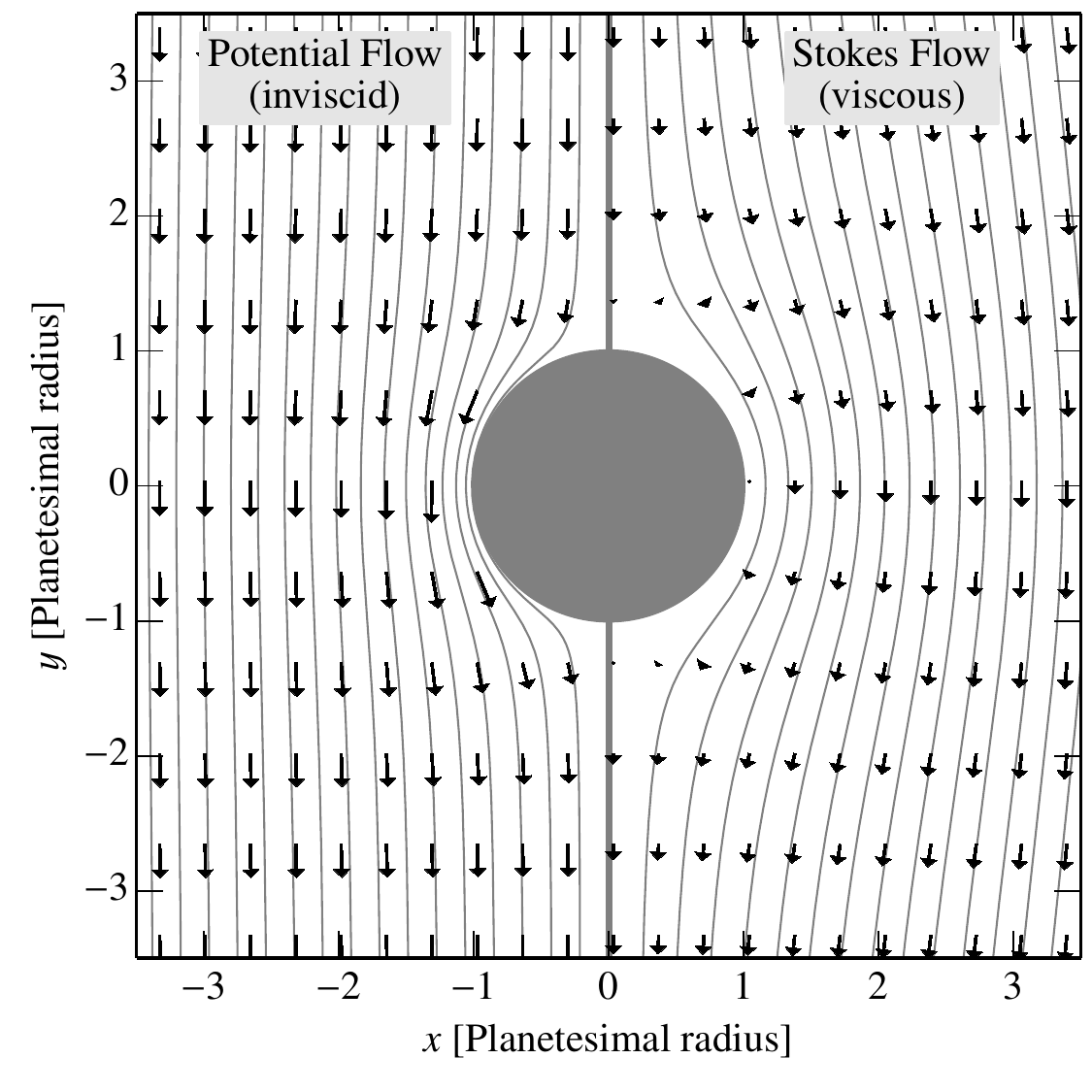}
    \caption{Velocity vectors and gas streamlines around a sphere (planetesimal) for potential flow (left half; \eq{Potential}) and Stokes flow (right half; \eq{Stokes}) as seen in the frame of the sphere.} %Note that for Stokes flow velocities close to the body slow down significantly (while those in the potential flow speed up modestly) and that it influences the flow at larger distances.}
    \label{fig:potstokes}
\end{figure}
We use two steady prescriptions for the gas flow in the vicinity of the planetesimals (see \fg{potstokes}): inviscid potential flow and (highly viscid) Stokes flow. The velocity of the potential flow past a sphere reads, in spherical coordinates \citep{Batchelor1967}:
\begin{equation}
    \mathbf{v}_\mathrm{Pot}  =   v_\mathrm{hw} \cos \theta \left (1 - \frac{R^3 }{r^3}\right )\mathbf{e}_r + v_{\mathrm{hw}} \sin \theta \left ( 1 + \frac{R^3 }{2r^3}\right )\mathbf{e}_\theta,
    \label{eq:Potential}
\end{equation}
with $\theta$ the angle between the direction of the unperturbed flow (here: $-\mathbf{e}_y$) and a point $\mathbf{x}$ in the local frame. Similarly, the Stokes flow solution reads \citep{Batchelor1967}:
\begin{equation}
    \mathbf{v}_\mathrm{Stk} =  v_\mathrm{hw} \cos \theta\left ( 1+\frac{R^{3}}{2r^{3}}-\frac{3R}{2r}  \right ) \mathbf{e}_r + v_\mathrm{hw}\sin \theta\left ( 1-\frac{R^{3}}{4r^{3}}-\frac{3R}{4r} \right )\mathbf{e}_\theta,
    \label{eq:Stokes}
\end{equation}
see \fg{potstokes}. Compared to Stokes flow, the potential solution does not contain \noeq{\propto}{1/r} terms. Stokes flow therefore modifies the unperturbed flow already at large distances from the planetesimal, which is a consequence of viscosity. Also, in Stokes flow the gas velocity is identically 0 at $R=0$, whereas in potential flow $v_r=0$ but $v_\theta \neq 0$ (see \fg{potstokes}).

To these solutions, we add the shear-corrected gas velocity, rendering the combined flow velocity of the gas to become:
\begin{equation}
    \mathbf{v}_g(x) =  \mathbf{v}_\mathrm{flow} - \frac{3}{2}\Omega_0 x \mathbf{e}_y,
    \label{eq:vgshear}
\end{equation} 
with $\mathbf{v}_\mathrm{flow}$ either the Stokes flow or the potential flow solution. Strictly, the addition in \eq{vgshear} is not self-consistent: it would, for example, violate the irrotational assumption used in deriving the potential flow. However, as the scales on which \eqs{Potential}{Stokes} and the Keplerian shear apply are very different ($R\Omega_0\ll v_\mathrm{hw}$), we expect the error to be negligible and irrelevant to the results of this work.

%Note that the shear term is negligible close to the planetesimal ($\sim R \Omega_0$) while the flow pattern is practically unperturbed for $x\gg R$. This allows us to add the shear to an irrotational flow and to neglect this complication.
\begin{figure}[t]
    \centering
    \includegraphics[width=9cm]{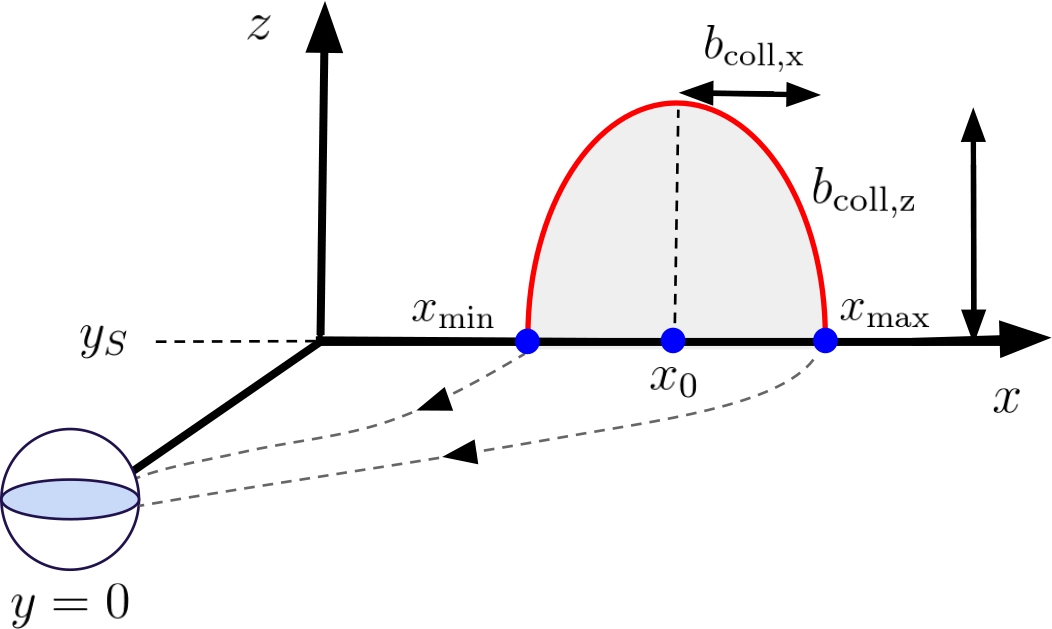}
    \caption{Sketch of the procedure to obtain the collisional cross section. In the comoving frame, the planetesimal resides at the origin ($y=0$) and pebbles start at a distance $y_\mathrm{S}$ far enough to be unperturbed by the gravity of the planetesimal. Integrations are first carried out in the $x$-direction and the impact parameter $b_\mathrm{coll,x}$ is determined from the first and last trajectory that hits the planetesimal ($x_\mathrm{min}$ and $x_\mathrm{max}$). Starting at the average of these two coordinates $x_0$, the integrations are extended to the $z$-direction to obtain the vertical impact parameter $b_\mathrm{coll,z}$.}
    \label{fig:sketch}
\end{figure}
The gas drag $\mathbf{F}_\mathrm{drag}$ felt by the particles reads:
\begin{equation}
    \mathbf{F}_\mathrm{drag} = 
    -\frac{\mathbf{v}_\mathrm{p} - \mathbf{v}_\mathrm{g}}{t_s},
\label{eq:drag}
\end{equation} 
where $\mathbf{v}_\mathrm{p}$ is the velocity of the particle, $\mathbf{v}_g$ the gas velocity and $t_s$ the stopping time for a spherical particle with radius $s$ \citep{Whipple1972,Weidenschilling1977}:
\begin{equation}
    t_s = 
    \left\{\begin{matrix}
    \displaystyle
    \frac{\rho_{\bullet\mathrm{s}} s}{\rho_{\mathrm{g}} c_{\mathrm{s}}} &                \textrm{Epstein regime: }  \quad s<\frac{9}{4}l_{\mathrm{mfp}} \\[5mm] 
    \displaystyle
    \frac{2 \rho_{\bullet\mathrm{s}}s^{2}}{9 \eta} & \textrm{Stokes regime:} \quad s \geq \frac{9}{4}l_{\mathrm{mfp}}\\ 
\end{matrix}\right.
\label{eq:tstop}
\end{equation} with $l_{\mathrm{mfp}}$ the mean free path of the gas molecules, $\rho_{\bullet,s}$ the internal particle density taken to be $1 \ \mathrm{g} \ \mathrm{cm^{-3}}$ and $\eta$ the dynamic viscosity. As our largest particle sizes are 30 $\mathrm{cm}$, these are the relevant stopping time regimes.% and the quadratic regime can be neglected.
\begin{table*}
\caption{Overview of the parameter space and the corresponding numerical results at given orbital distance $r$.}             
\label{tab:results}      
\centering          
\begin{tabular}{l l l l l l l l l }     % 7 columns 
\hline
\hline
$r$ & Flow & $s$ & $\tau_s$ & $f_\mathrm{coll,min}$ & $t_\mathrm{gr,max}$ & $R_\mathrm{gr,max}$ & $R_\mathrm{PA}$ & comments\\
    &   & [cm]  &   &   &   [yr]    & [km]  & [km] \\
\hline                    
   1 AU &   pot & 0.01  & $8.5\times   10^{-6}$ & 0.001 & $1.8 \times 10^9$ & 40 & 50 & \\ 
        &   stk & 0.01  & $8.5\times 10^{-6}$ & 0.005 & $6.9 \times 10^7$& 10 & 20& \\  
        &   pot & 0.03  & $2.5\times 10^{-5}$ & 0.03 & $1.4 \times 10^8$ & 80 & 90 & \\
        &   stk & 0.03  & $2.5\ \times 10^{-5}$ &0.03  & $2.4 \times 10^7$ & 30 & 40 & \\
        &   pot & 0.1   & $8.5\ \times 10^{-5}$ & 0.3  & $1.7 \times 10^7$ & 130 & 150 & \\
        &   stk & 0.1   & $8.5\ \times 10^{-5}$ & 0.2  & $7.7 \times 10^6$ & 50 & 50 & \\
        &   pot & 0.3   & $2.5\ \times 10^{-4}$ & 0.8 & $5.3 \times 10^6$ & 110 & 130 & \\
        &   stk & 0.3   & $2.5 \ \times 10^{-4}$ & 0.5 & $3.3 \times 10^6$ & 50 & 60 & \\
        &   pot & 1     & $8.5\ \times 10^{-4}$ & 1.0 & $2.9 \times 10^6$ & 90 & 110 & \\
        &   stk & 1     & $8.5\ \times 10^{-4}$ & 0.8 & $2.7 \times 10^6$ & 40 & 50 & \\
        &   pot & 3     & $2.5\ \times 10^{-3}$ & 1.0 & $2.4 \times 10^6$ & 80 & 90 & \\
        &   stk & 3     & $2.5\ \times 10^{-3}$ & 1.0 & $2.4 \times 10^6$ & 70 & 80 & \\
        &   pot & 10    & $2.5\ \times 10^{-2}$ & 1.0 & $2.2 \times 10^6$ & 80 & 150 & \\
        &   pot & 30    & $2.3\ \times 10^{-1}$ & 1.0 & $2.2 \times 10^6$ & 80 & 300 & \\
\hline                    
   3 AU &   pot & 0.01  & $4.4\ \times 10^{-5}$ & 0.75 & $1.2 \times 10^8$ & 110 & 120 & \\
        &   pot & 0.03  & $1.3\ \times 10^{-4}$ & 1.0 & $6.5 \times 10^7$ & 90 & 110 & \\
        &   pot & 0.1   & $4.4\ \times 10^{-4}$  & 1.0 & $5.0 \times 10^7$ & 80 & 130 & \\
        &   pot & 0.3   & $1.3\ \times 10^{-3}$  & 1.0 & $4.6 \times 10^7$ & 80 & 150 & \\
        &   pot & 1     & $4.4\ \times 10^{-3}$  & 1.0 & $4.5 \times 10^7$ & 80 & 200 & \\
        &   pot & 3     & $1.3\ \times 10^{-2}$  & 1.0 & $4.4 \times 10^7$ & 80 & 250 & \\
        &   pot & 10    & $4.4\ \times 10^{-2}$  & 1.0 & $4.4 \times 10^7$ & 80 & 350 & \\
        &   pot & 30    & $5.7\ \times 10^{-2}$  & 1.0 & $4.4 \times 10^7$ & 50 & 200 & $3\rho_\mathrm{g}$, see \fg{diffmodels}\\
        &   pot & 30    & $1.3\ \times 10^{-1}$  & 1.0 & $4.4 \times 10^7$ & 80 & 450 & \\
        &   pot & 30    & $1.3\ \times 10^{-1}$  & 1.0 & $4.4 \times 10^7$ & 30 & 150 & $T = 90 \ \mathrm{K}$, see \fg{diffmodels}\\
        
\hline    
   10 AU &   pot & 0.01  & $2.7\ \times 10^{-4}$ & 1.0 & $1.3 \times 10^7$ & 80 & 150 & \\
        &   pot & 0.03  & $8.1\ \times 10^{-4}$  & 1.0 & $1.3 \times 10^7$ & 70 & 200 & \\
        &   pot & 0.1   & $2.7\ \times 10^{-3}$  & 1.0 & $1.2 \times 10^7$ & 80 & 250 & \\
        &   pot & 0.3   & $8.1\ \times 10^{-3}$  & 1.0 & $1.2 \times 10^7$ & 80 & 350 & \\
        &   pot & 1     & $2.7\ \times 10^{-2}$  & 1.0 & $1.2 \times 10^7$ & 80 & 500 & \\
        &   pot & 3     & $8.1\ \times 10^{-2}$  & 1.0 & $1.2 \times 10^7$ & 80 & 600 & \\
        &   pot & 10    & $2.7\ \times 10^{-1}$  & 1.0 & $1.2 \times 10^7$ & 80 & 750 & \\
        
\hline                  
\end{tabular}

\tablefoot{Columns denote: the orbital disk radius $r$, the adopted flow solution (potential or Stokes), the particle radius $s$, the dimensionless stopping time $\tau_s=t_s\Omega$, the minimal collision factor $f_\mathrm{coll, min}$ obtained from the results shown in \fg{fcoll}, the maximum growth timescale $t_\mathrm{gr,max}$ obtained from the results shown in \fg{Tgrowth} (unless denoted otherwise in the comment section), the corresponding planetesimal radius $R_\mathrm{gr,max}$ and the planetesimal radius where pebble accretion starts $R_\mathrm{PA}$.}
\end{table*}
\begin{figure*}
    \includegraphics[height=7cm]{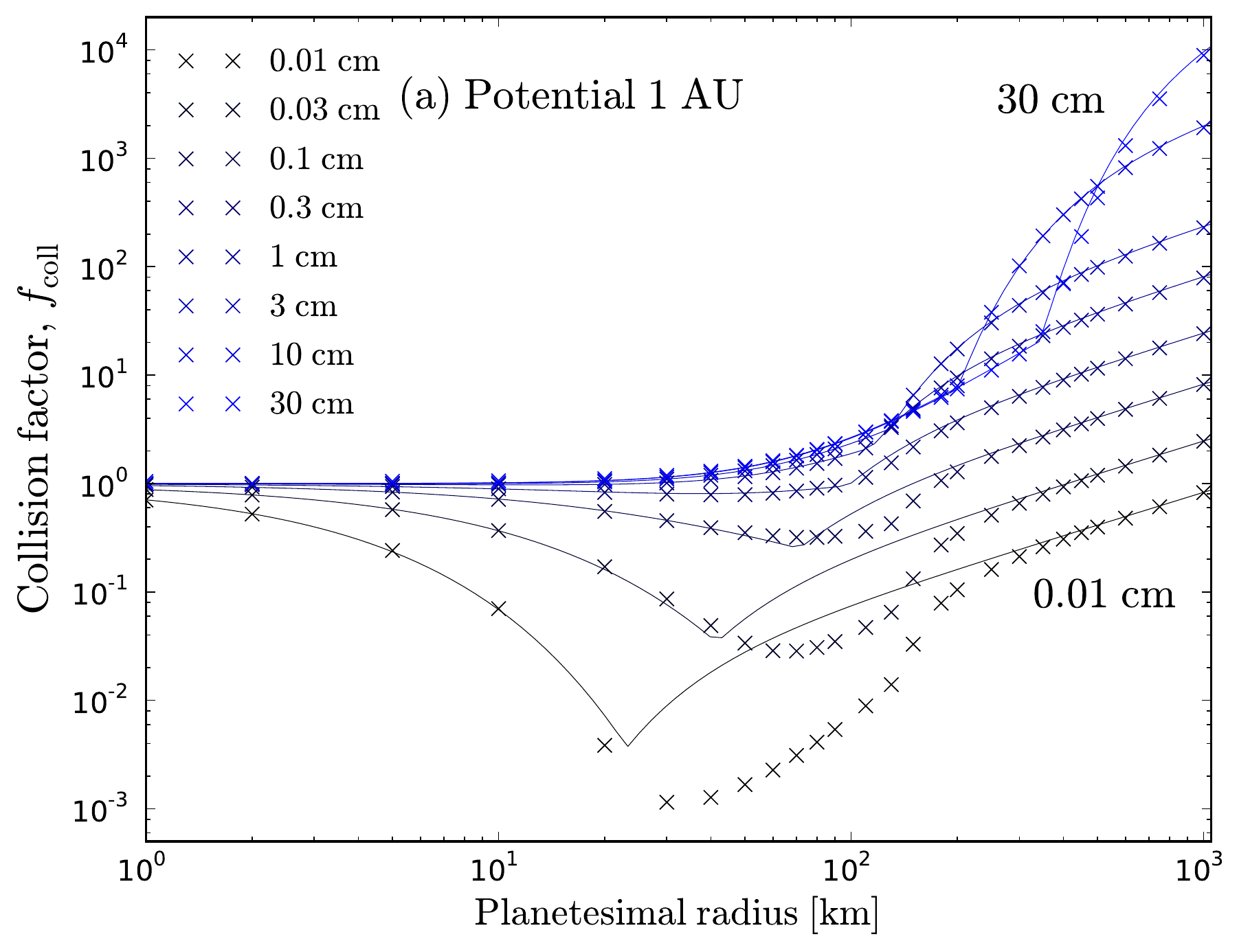}
    \includegraphics[height=7cm]{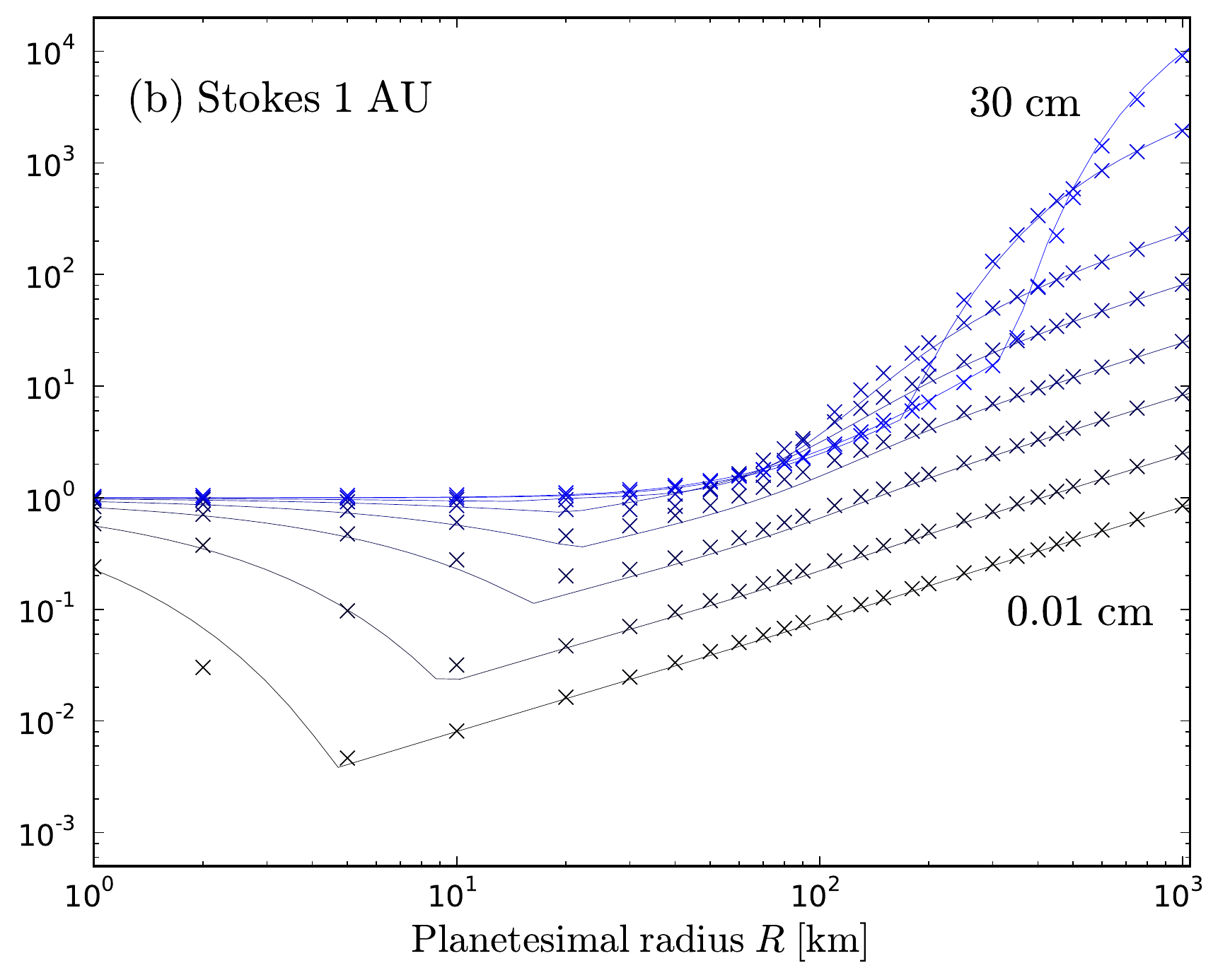} \\ 
    \includegraphics[height=7cm]{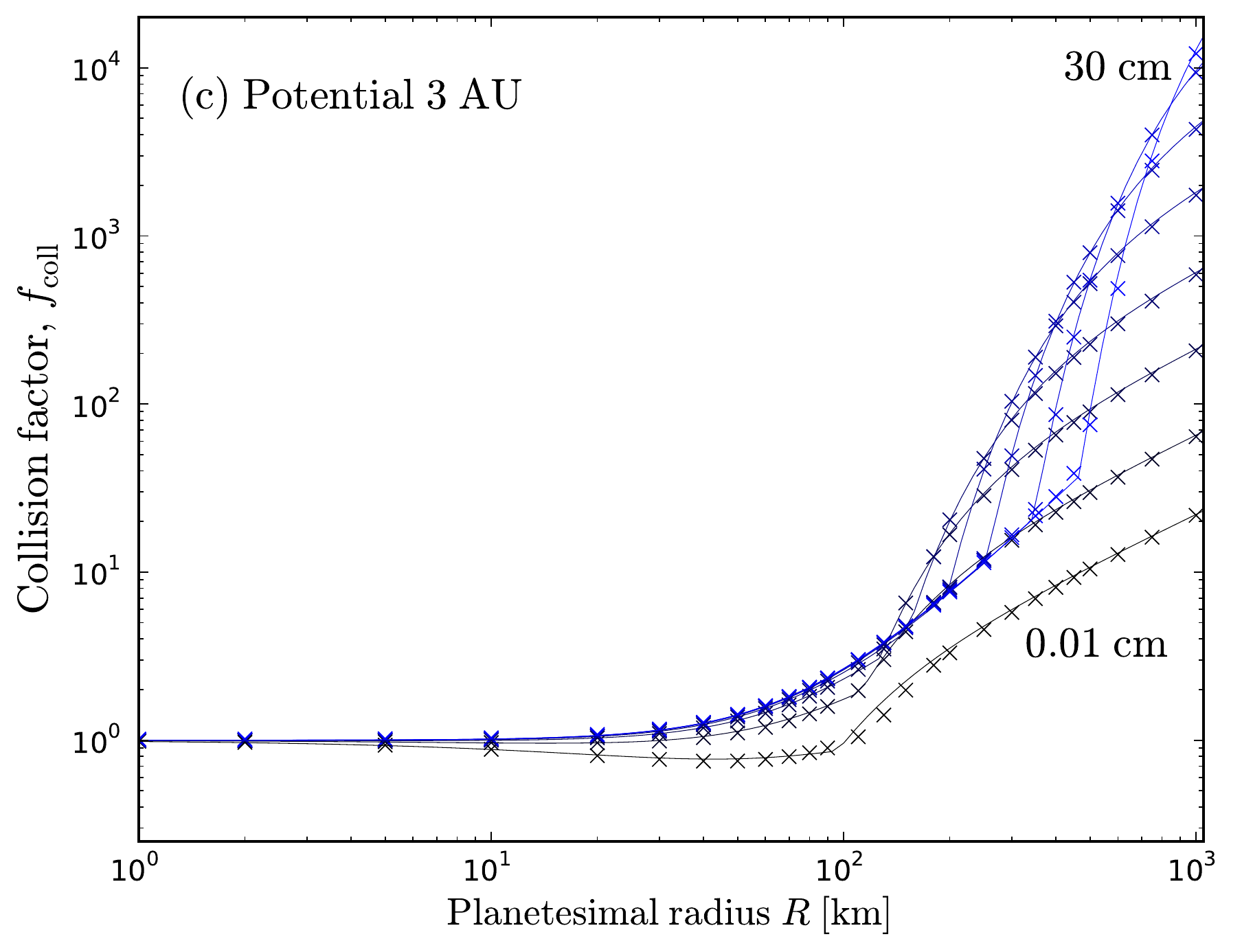}
    \includegraphics[height=7cm]{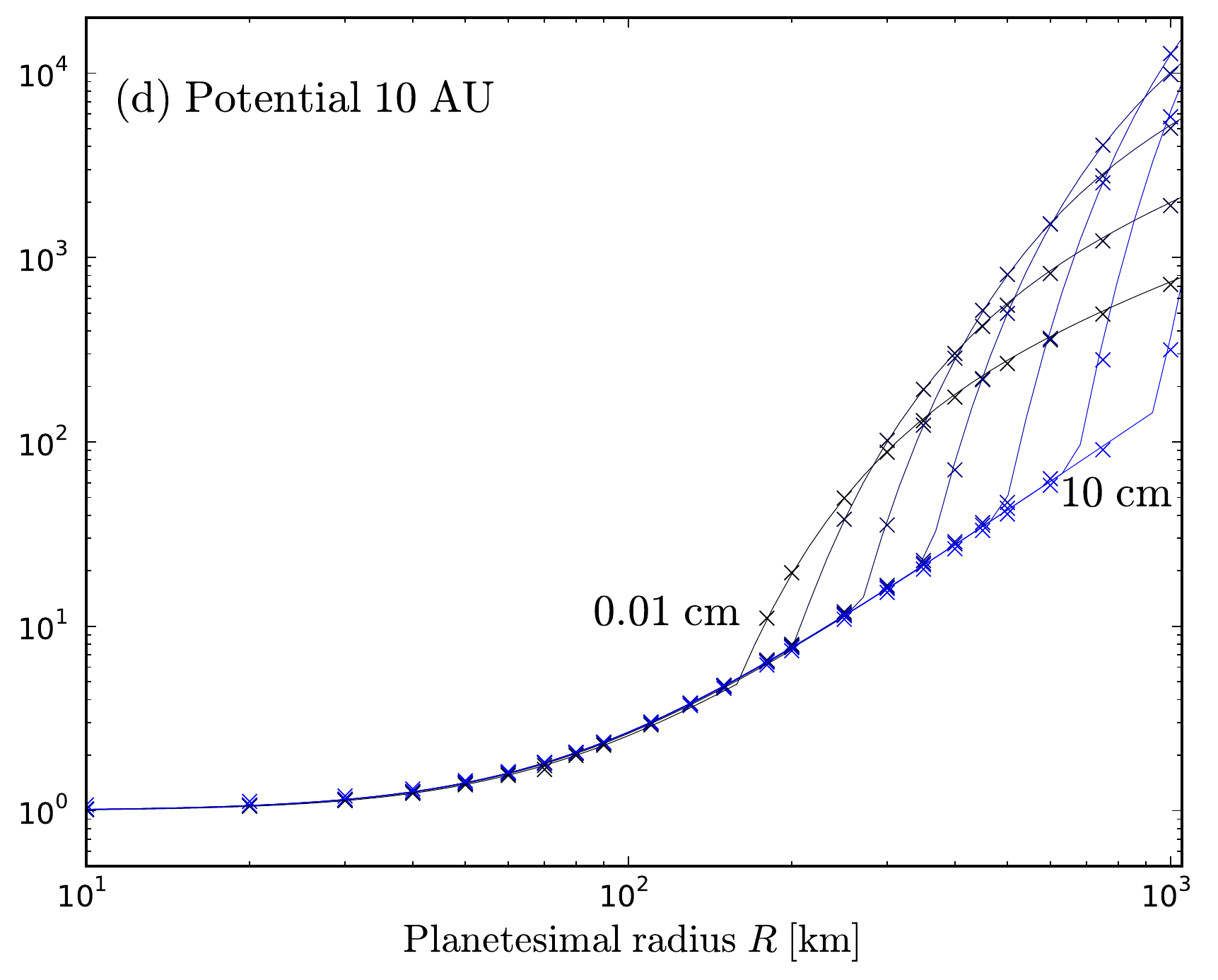}
    \caption{Collision factor $f_\mathrm{coll}$ as function of planetesimal radius $R$ for a constant internal density $\rho_\bullet = 1 \ \mathrm{g \ cm^{-3}}$. Crosses indicate numerical data while solid lines are fits. The color scale ranges from dark to light blue for increasing particle radius. (a) Results obtained with the (inviscid) potential solution (\Eq{Potential}) at $1 \ \mathrm{AU}$ orbital distance from the star. For small planetesimal size, the collision factor is approximately geometrical ($f_\mathrm{coll} \approx 1$). However, for small pebble sizes, aerodynamical deflection results in a significant drop of $f_\mathrm{coll}$  (see \Tb{results} for the minimum collision efficiency, $f_{\mathrm{coll,min}}$). (b) Results with the (highly viscid) Stokes flow solution (\Eq{Stokes}). Settling occurs at smaller sizes due to the increase in encounter time. (c) Results at $3 \ \mathrm{AU}$ orbital distance show a shallow barrier only for the lowest particle size. (d) At 10 AU a barrier is no longer present, because of the reduced gas density.}
    \label{fig:fcoll}
\end{figure*}

\begin{figure*}
    \includegraphics[height=7cm]{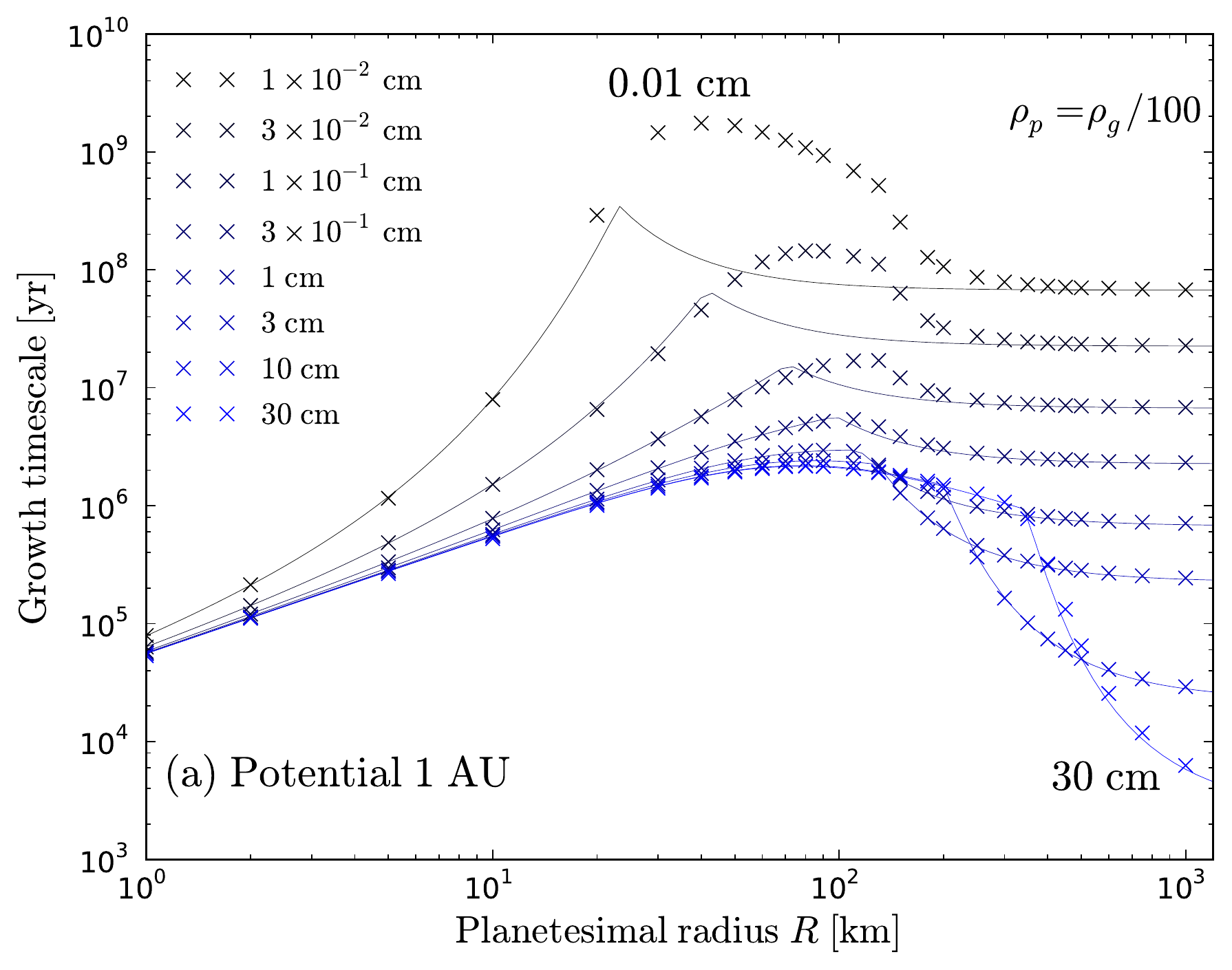}
    \includegraphics[height=7cm]{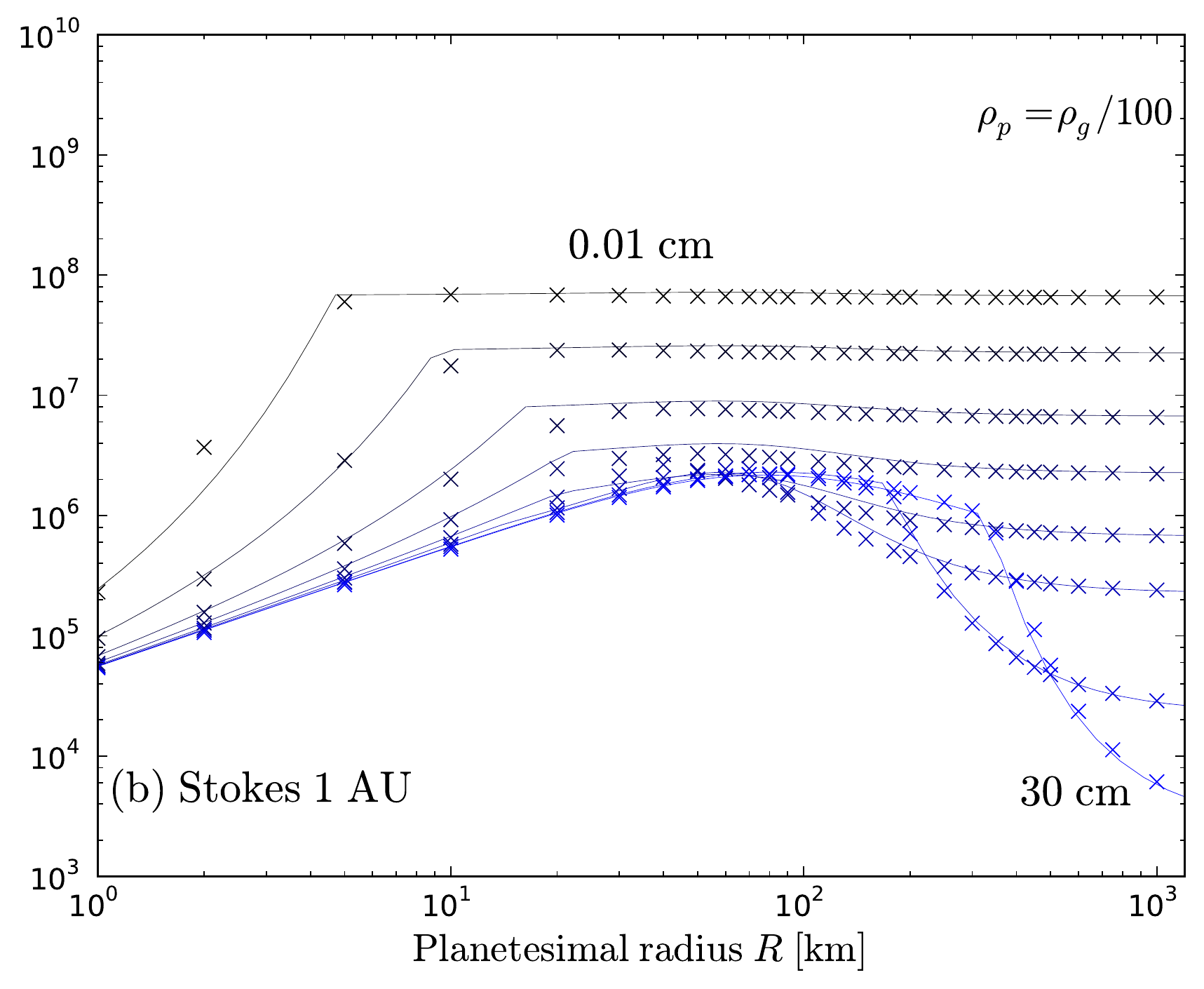} \\ 
    \includegraphics[height=7cm]{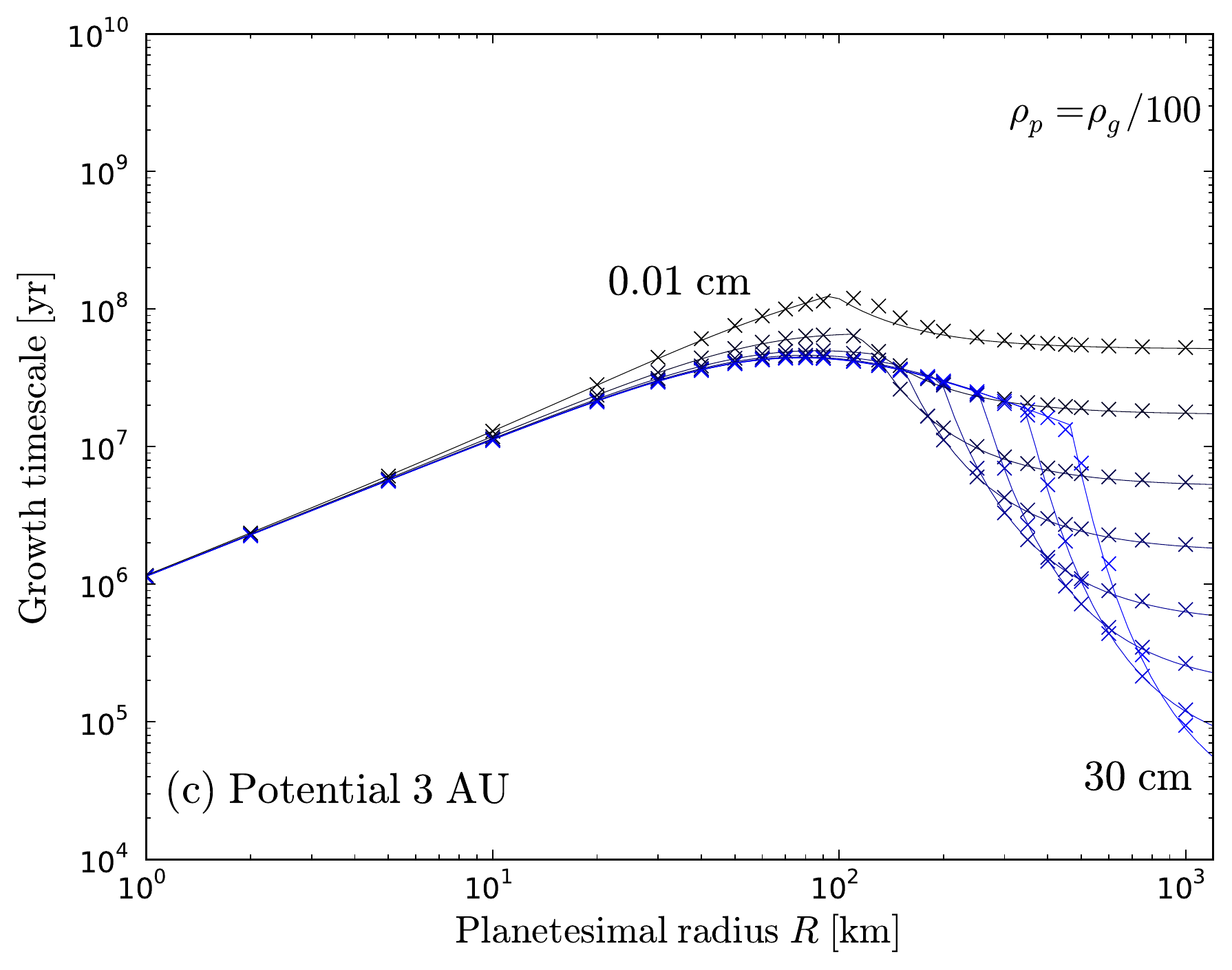}
    \includegraphics[height=7cm]{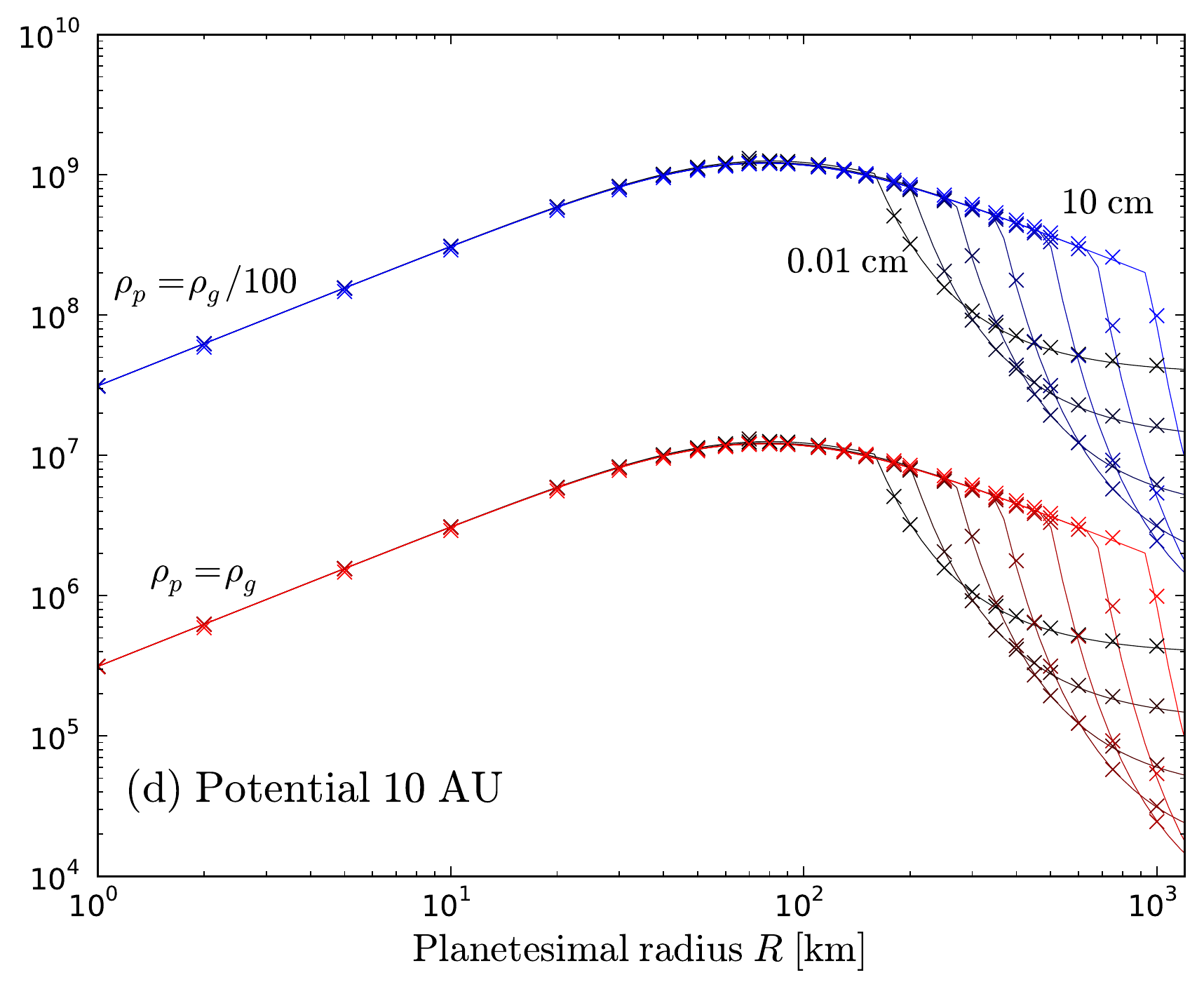}
    \caption{Growth timescale $t_\mathrm{growth}$ as function of planetesimal radius $R$ for a constant internal density $\rho_\bullet = 1 \ \mathrm{g \ cm^{-3}}$. Crosses indicate numerical data while the solid lines are fits. The color scale goes from dark blue scale to light blue scale transitioning from lowest particle radius to highest particle radius respectively. (a) $t_\mathrm{growth}$ obtained with the potential flow solution at $1 \ \mathrm{AU}$ orbital distance and a pebble density of $\rho_p = \rho_{g} / 100$, showing maxima's at $R_\mathrm{gr,max} \sim 100 \ \mathrm{km}$ (b) $t_\mathrm{growth}$ with the Stokes flow solution at $1 \ \mathrm{AU}$ orbital distance and a pebble density of $\rho_p = \rho_{g} / 100$ , showing early settling and flattening out of the curves. (c) $t_\mathrm{growth}$ with the potential flow solution at $3 \ \mathrm{AU}$ orbital distance and a pebble density of $\rho_p = \rho_{g} / 100$. (d) $t_\mathrm{growth}$ with the potential flow solution at $10 \ \mathrm{AU}$ orbital distance and a pebble density of $\rho_p = \rho_{g} / 100$ (top, blue) and a pebble density of $\rho_p = \rho_{g}$ (bottom, red).}
    \label{fig:Tgrowth}
\end{figure*}

\subsection{Numerical integrations}
We numerically integrate the particle streamlines with the Runge-Kutta-Fehlberg variable step method (RKF45). \changed{This integration method uses variable steps and compares a fourth order solution with a fifth order solution to determine the next step size \citep{Fehlberg1969,eshagh2005,OrmelKlahr2010}. We use a relative error tolerance of $10^{-8}$, small enough to ensure convergence.} The goal of the integrations is to determine the collisional cross-sections of pebbles of radius $s$ hitting a planetesimal of radius $R$. This is achieved by integration of the pebble equation of motion, \Eq{eqofmotion}, starting at a distance $y_S$ from the planetesimal, see \fg{sketch}. The value of $y_S$ has been chosen large enough to ensure that at this distance the unperturbed flow solution holds \citep{Weidenschilling1977}:
\begin{equation}
    v_x = v_{x,\infty} =-\frac{2v_\mathrm{hw}\tau_s}{1 + \tau_s^2},
    \label{eq:vxin}
\end{equation}
\begin{equation}
    v_y(x_S) = v_{y,\infty} -\frac{3}{2}\Omega_0 x_\mathrm{s} 
    = -\frac{v_\mathrm{hw}}{1 + \tau_s^2} -\frac{3}{2}\Omega_0x_\mathrm{S},
    \label{eq:vyin}
\end{equation}
\changed{with $v_{x,\infty}, v_{y,\infty}$ the (unperturbed) radial and azimuthal drift velocities of the pebble,} $-\frac{3}{2}\Omega_0x_\mathrm{S}$ the correction for Keplerian shear, $x_S$ the starting $x$-coordinate of the particles and $\tau_s$ the dimensionless stopping time $\tau_s = \Omega_0 t_s$.

The initial `starting line' $y=y_S$ is chosen far enough far enough from to assure that the gravity of the planetesimal is small in comparison to the gas drag force. Specifically, we adopt:
\begin{equation}
    y_S = \max \left( 50R, 10^3\sqrt{\frac{GM}{v_\mathrm{hw}/t_s}} \right).
\end{equation}
Starting in the midplane ($z=0$) we first vary $x_\mathrm{S}$ and determine whether the integration results in a hit or a miss. Thus, we search for the positions $\{x_\mathrm{min}$, $x_\mathrm{max}\}$ separating hits from misses. \changed{The condition for a hit is given by: 
\begin{equation}
    \sqrt{x^2 + y^2 + z^2} - R \leq  0.
\end{equation}} 
%We ensure a precision of at least 1\% and 
We then determine the impact parameter as $b_\mathrm{coll,x} = (x_\mathrm{max} - x_\mathrm{min})/2$. %\label{eq:bcollx},
%\end{equation}
By focusing our search around $x=x_\mathrm{min},\ x_\mathrm{max}$ we ensure that the error in $b_\mathrm{coll,x}$ is at most 1\%. Next, we integrated in the $z$-direction starting at the midpoint $(x_0 = (x_\mathrm{max} + x_\mathrm{min})/2, y_S, z=0)$ to determine $b_\mathrm{coll,z}$ (see \fg{sketch}).
%\begin{equation}
%    b_\mathrm{coll,z} = z_\mathrm{max},
%\label{eq:bcollz}
%\end{equation}
We assume that the impact geometry is ellipsoidal and that the impact area (at $y=y_S$) is $A=\pi b_\mathrm{coll,x} b_\mathrm{coll,z}$.  \changed{The corresponding pebble flux is $F = A \mathbf{\hat{n}} \cdot \mathbf{v} = \pi b_\mathrm{coll,x}b_\mathrm{coll,z} v_y(x_0)$, with $\mathbf{\hat{n}}$ the unit vector perpendicular to $A$. Multiplied by $\rho_p$, the spatial density of pebbles in the midplane, this gives the total accretion rate $\dot{M}$. As in the geometric limit the flux is $\pi R^2 v_\mathrm{hw}$, the collisional efficiency is therefore}:
\begin{equation}
    f_\mathrm{coll}
    = \frac{\pi b_\mathrm{coll,x} b_\mathrm{coll,z} v_y(x_0)}{\pi R^2 v_\mathrm{hw}}.
%    \approx \pi b_\mathrm{eff}^2 v_\mathrm{hw} 
    \label{eq:fcoll}
\end{equation}
In terms of $f_\mathrm{coll}$ the accretion rate is:
\begin{equation}
    \dot{M} = \pi R^2 v_\mathrm{hw}\rho_\mathrm{p} f_\mathrm{coll}.
    \label{eq:massroc}
\end{equation}
%The mass accretion rate for a planetesimal embedded in a sea of pebbles is given by $f_\mathrm{coll}$  multiplied by the geometrical flux and the pebble density at the midplane:

%\comm{Do we still use this??}However, because of Keplerian shear $b_\mathrm{coll,x}$ depends on $x_0$ (and indirectly on $y_S$). However, continuity requires that the product We therefore define an effective impact parameter $b_\mathrm{eff}$:

%with $b_\mathrm{eff}$ the impact parameter corrected for shear, which is independent of the arbitrarily chosen starting point $y_S$. \Eq{flux} thus provides $b_\mathrm{eff}$ from which we define the effective collision factor as:
%\begin{equation}
%    f_\mathrm{coll} \equiv  \frac{\pi b_\mathrm{eff}^2}{\pi R^2}=\left (\frac{b_\mathrm{eff}}{R}  \right )^2,
%    \label{eq:fcoll}
%\end{equation}

\section{Results}
\label{sec:Results}
Results for $f_\mathrm{coll}$ are obtained as function of particle radius $s$, planetesimal radius $R$, disk orbital radius (1, 3, and 10 AU) and flow pattern (Stokes or potential). We adopt internal densities of $\rho_\bullet = 1 \ \mathrm{g} \ \mathrm{cm}^{-3}$ for both planetesimals and pebbles. Results of the integrations are given in \Tb{results} and in \fgs{fcoll}{Tgrowth}. In the figures, crosses indicate the numerical results while solid lines represent fits (see \se{analysis}).

\subsection{Collision efficiencies}
\label{sec:CollEff}
For small planetesimals, the collisional cross section equals the geometric cross section, $f_\mathrm{coll} \approx 1$. However, the Stokes number (\eq{Stnr}) decreases for increasing planetesimal sizes. When it reaches unity, aerodynamic deflection can become important as the particle will adjust to the flow on a timescale ($t_s$) that is shorter than the crossing time $R/\Delta v$. Furthermore, for small particles, (below 0.1 $\mathrm{cm}$) the 2-body gravitational force is weak. Consequently, the collision factor becomes (much) smaller than unity. This is consistent with the non-gravitational hydrodynamical simulations by \citet{SekiyaTakeda2003}; under these conditions, particles hardly hit the planetesimal. For our smallest size of $10^{-2}$ cm, collisional deflection factors reach a minimum of $10^{-3}$ for the potential case and slightly higher for the Stokes case. In addition, in Stokes flow the minimum occurs at smaller planetesimals. On the other hand, it can be seen from  \fg{fcoll}(a) and (b) that $f_\mathrm{coll}$ never reaches zero. The minimum value for the collision factor, $f_\mathrm{coll, min}$, is listed in \Tb{results}.  

This difference between the two flow patterns can be explained as follows. In both cases the radial velocity is zero at the surface of the planetesimal. However, where the angular component is zero in the viscous case (zero slip boundary), it is on the order of the headwind velocity for potential flow. For the potential case, this explains the deep\? barrier that is seen in \fg{fcoll}a: particles feel a centrifugal force (of order \noeq{\sim}{v_\mathrm{hw}^2/R}) that competes with the gravitational attraction. On the other hand, for the Stokes case the almost vanishing azimuthal velocities means that the crossing times become significantly longer than $R/v_\mathrm{hw}$, which has two consequences: (i) aerodynamic deflection becomes already effective at higher Stokes number (smaller planetesimal size for fixed $t_s$); and (ii) gravity has more time to act, promoting accretion by settling.  As a result, the minimum of the collisional efficiencies for Stokes flow occurs at a smaller planetesimal radius.

From \Tb{results} two modes can be identified. (i) For pebbles smaller than $0.1 \ \mathrm{cm}$ hydro-effects operate to slow down particles within the time pebbles encounter the planetesimal. The collision factor steadily decreases, with $f_\mathrm{coll,min}$ possibly $\ll$1 due to aerodynamic deflection, before settling interactions reverse the decline. (ii) Larger pebbles  ($s>0.1\ \mathrm{cm}$ at 0.1 AU) never experiences aerodynamic deflection. Instead, we see a gradual \textit{increase} of $f_\mathrm{coll}$, characteristic of the smooth transition to Safronov gravitational focusing. Collision factors then increase with planetesimal size. Eventually, gravity becomes strong enough to trap pebbles in the Hill sphere, initiating the onset of pebble accretion, which expresses itself as a steep upturn\? in $f_\mathrm{coll}$. 

As was explained above, the onset of pebble accretion also depend on disk radius.%due to the increased crossing time, pebble accretion is initiated at smaller planetesimal sizes in the Stokes flow solution. However, when $f_\mathrm{coll}\gg1$ collision efficiency factors become independent of the nature of the flow around the objects.
The integrations at $3 \ \mathrm{AU}$ and $10 \ \mathrm{AU}$ with the potential flow solution are shown in \fg{fcoll}c and \fg{fcoll}d. In \fg{fcoll}c only the smallest particle shows a barrier (see \Tb{results}). For larger particle radii, the collision efficiency factors starts in the geometric regime, smoothly transition to the Safronov regime, before entering the settling regime. Since stopping times increase with disk radius and with particle radius, the onset of pebble accretion occurs at larger planetesimal radius, explaining the prolonged duration of ballistic encounters.
%In \fg{fcoll}d no barrier is present anymore and the onset of pebble accretion is graphically visible for all curves, characterized by the sharp transition from the Safronov regime

\subsection{The growth timescale for sweepup}
\Eq{massroc} defines the timescale for the planetesimal to $e$-fold its mass:
\begin{equation}
t_\mathrm{growth} = \frac{M}{\dot{M}} =\frac{ 4 \rho_\bullet R}{3 v_\mathrm{hw} \rho_p f_\mathrm{coll}}.
\label{tgrowth}
\end{equation}

The growth timescales for the same parameter space as our results for $f_\mathrm{coll}$ are shown in \fg{Tgrowth} and listed in \Tb{results}. Also, the crosses denote numerical data-points and solid lines are fits (see \se{analysis}). The density of pebbles is taken to equal $\rho_p = \rho_g/ 100$, the typical solid-to-gas ratio of the gas\changed{, except in \fg{Tgrowth}(d) where we also consider the case of equal pebble and gas density}. \fg{Tgrowth}(a) and \fg{Tgrowth}(b) show growth timescales using the potential and Stokes flow solutions at $1 \ \mathrm{AU}$ orbital distance. The time needed for the planetesimal to increase its mass by a factor $e$, increases for small planetesimal size $R$ up to a maximum $R_\mathrm{gr,max}$. In the potential case, this maximum is centered around $R_\mathrm{gr,max} \sim 100 \ \mathrm{km}$ (see \Tb{results}). For the simulations using the Stokes flow pattern, the maximum occurs at smaller pebble sizes and evaluates to shorter times.

The features present in \fg{Tgrowth} reflect those of \fg{fcoll}. In the geometrical regime, growth timescales increase since the dependency of planetesimal mass on radius is cubic and the dependency on the impact parameter is quadratic. For pebble sizes \noeq{\lesssim}{0.1} $\mathrm{cm}$, the increase in growth time is especially dramatic due to the aerodynamical deflection discussed in \se{CollEff}. Pebbles \noeq{\gtrsim}{10\ \mathrm{cm}} never experience aerodynamic deflection, but instead enter the Safronov regime (as was discussed in \se{CollEff}). Although Safronov focusing causes a decrease in growth timescale with planetesimal radius (runaway growth), growth timescales remain long. Nevertheless, growth inevitably transitions to the pebble accretion regime, which commences with a steep drop in $t_\mathrm{growth}$.

At 3 and 10 AU orbital distances hydrodynamical deflection is absent, except for pebbles smaller than $0.01 \ \mathrm{cm}$ at 3 AU. For larger pebbles, the growth time initially decreases slowly (Safronov), but is followed by a sharp transition to pebble accretion  %In this case, growth timescales are given by geometrical considerations until gravitational focusing becomes important 
(\fg{Tgrowth}(c) and \fg{Tgrowth}(d)). Because gas densities decrease with disk radius (\Eq{density}), $t_\mathrm{growth}$ increases for most pebble radii.
We see that in many cases peak growth timescales are much longer than the expected lifetimes of 6-10 Myr of gas-rich protoplanetary disks around low to intermediate mass stars \citep{Yasui2012,Pfalzner2014,Ribas2015}. Planet formation by pebble sweep-up starting off from small planetesimals would therefore be too slow.
However, particle settling to the mid-plane of the disk\citep{Youdin2007,Andrews2009} will increase pebble densities. \changed{In \fg{Tgrowth}(d) we have included the case that $\rho_g = \rho_p$, which reduces the timescales a hundred fold. Nevertheless, at 10 AU, the resulting peak growth timescales of $10^7$ yr are still rather long.}
%All the curves again sh%ow that the maximum of $t_\mathrm{growth}$ is centered at $R_\mathrm{gr,max} \sim 100 \ \mathrm{km}$ (see \Tb{results}). 

\subsection{Varying the disk model}
\label{sec:disk-model}
Until now we have adopted the disk profile as given by \eqs{Temp}{surfdens}. In reality, however, there is no clear physical reason why protoplanetary disk should be characterized by these MMSN-like profiles. We therefore consider the effect of changing the temperature and density.
%Another point\? is that the surface density profile of our fiducial (MMSN) perhaps results in too low densities in the outer regions of the disk \rico{and temperature too high?}.
In \fg{diffmodels} we show in green the growth timescale of a 30 cm particle at 3 AU for a disk with a temperature of 90 K that is half that of the standard model (black). Such colder disk conditions may follow from more efficient cooling \citep{BitschEtal2015}. The key consequence of the temperature drop is that it reduces the headwind of the disk (\eq{headwind}) by a factor 2. The reduced headwind increases the interaction timescale, providing more time for gravity to act. Thus, settling interactions now occur at a small planetesimal size (200 km instead of 450 km) and the peak growth time also shift to smaller planetesimal sizes (\noeq{\simeq}{50} km) because Safronov focusing also profits from a lower $v_\mathrm{hw}$. On the other hand, geometric encounters, which scale \noeq{\propto}{v_\mathrm{hw}}, are less effective, explaining the increase of the growth timescale at small $R$. Consequently the peak growth timescale hardly changes; the growth time bottleneck remains but shift to smaller sizes.

The green line in \fg{diffmodels} corresponds to the case of a more massive disk compared to the default model. Here, we have adopted a $-0.5$ exponent in the surface density expression of \eq{surfdens}, which results in an increase of the gas density by a factor 3. The pebble density $\rho_p$ is, however, kept the same as in the default model. Because of the shallower density profile, the headwind velocity drops slightly (to 40 $\mathrm{m\ s^{-1}}$), but the main effect of the enhanced gas density is that the 30 cm size particles have become aerodynamically smaller (lower $t_s$). Increasing the pebble density by a factor of three such that $\rho_g/\rho_p$ is back at 100 (not shown) causes the growth timescale to drop by a factor of three.

%exponent in \eq{surfdens} is $-0.5$ instead of $-1.5$. The change in the power-law index introduces a number of changes to the disk structure, which effects we consecutively show in \fg{diffmodels}. \changed{}First it reduces the headwind velocity to 40 $\mathrm{m \ s^{-1}}$. This increases the growth time in the geometrical regime since the flux decreases (see \eq{massroc}). On the other hand, a lower $v_\mathrm{hw}$ increases the interaction time for pebbles, which shifts $R_\mathrm{PA}$ to smaller planetesimal size. As a result, the reduction of the headwind does not have a clear effect on the peak growth timescale.  Second, a three fold increase in gas density renders particles aerodynamically smaller . Although this does not affect ballistic encounters (geometrical and Safronov regimes), it again slightly decreases the size where pebble accretion starts.
%. significant drop in peak growth timescales because of smaller stopping times in combination with a higher pebble flux.

\begin{figure}[t]
    \sidecaption
    \includegraphics[width=9cm]{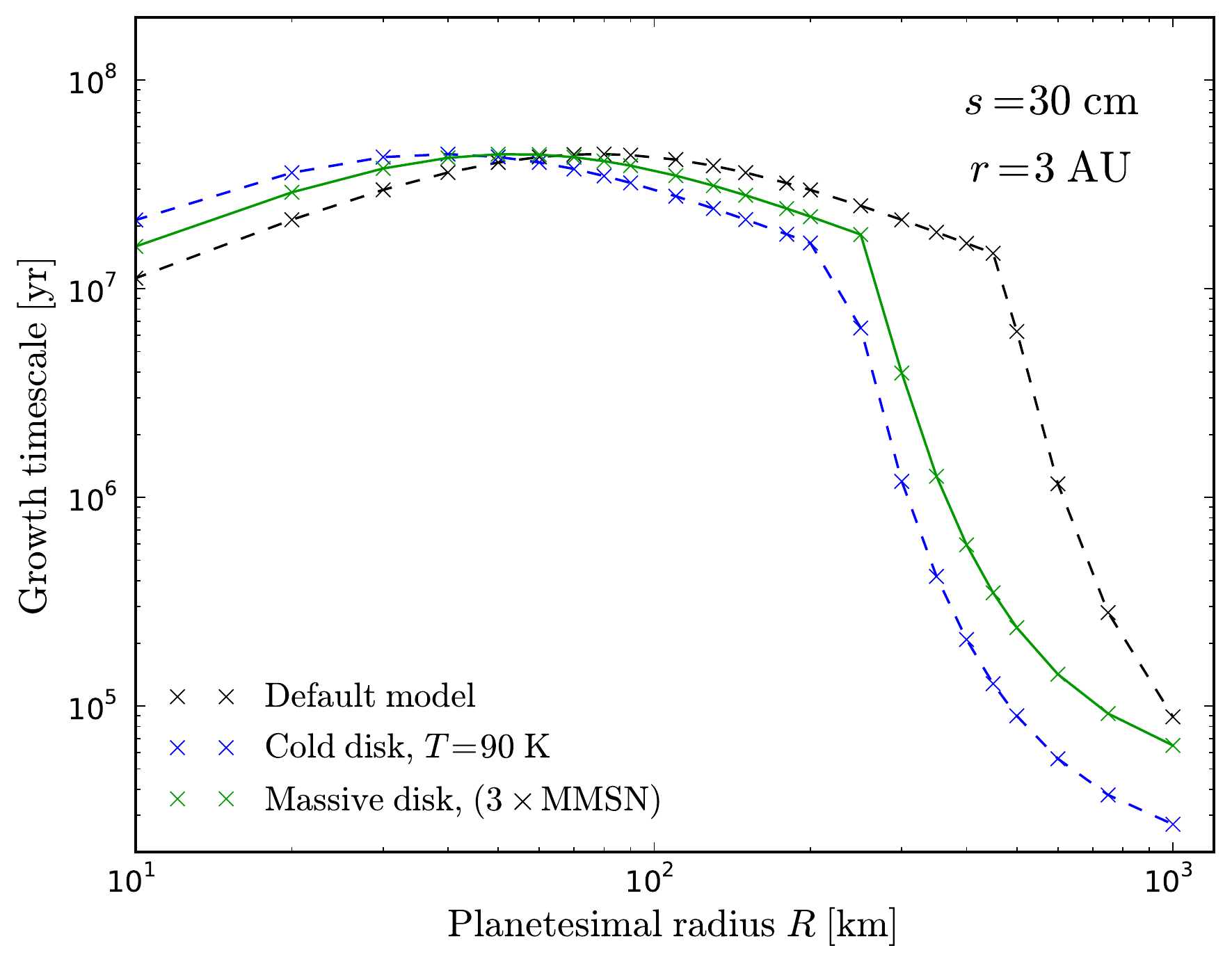}
    \caption{Growth timescale $t_\mathrm{growth}$ for a pebble of radius s = 30 cm as function of planetesimal radius $R$ at 3 AU. The black dashed curve shows the default model (as in \fg{Tgrowth}). The other curves consecutively give the effects of: \changed{a colder disk resulting in a headwind of $v_\mathrm{hw} = 28 \
\mathrm{cm \ s^{-1}}$ (blue dashed line) and a three fold increase in gas density (green curve) with a $-0.5$ power-law index for the surface density (\eq{surfdens})}. The crosses represent numerical results and the solid (and dashed) lines represent fits.}
    \label{fig:diffmodels}
\end{figure}
\section{Analysis: the onset of pebble accretion}
\label{sec:analysis}
\subsection{Fitting expressions for ballistic and settling interactions}
Pebble accretion occurs when particles settle to the planetesimal at terminal velocities $v_\mathrm{settl}=f_g t_s$. The equation that determines the impact parameter $b_\mathrm{set}$ of these settling encounters is the cubic
\begin{equation}
    b^2 \left(v_\mathrm{hw} +\frac{3}{2}\Omega b\right)
    = 4 GM_p t_s
    \label{eq:cubic}
\end{equation}
\citep{OrmelKlahr2010}. This equation follows from equating the settling timescale $b/v_\mathrm{settl}$ with the encounter timescale, $b/\Delta v$ where $\Delta v$, the approach velocity, is the term in braces in \eq{cubic}. The numerical factor 4 follows from the strongly coupled limit ($t_s \rightarrow 0$), which allows closed-form analytical solutions \citep{OrmelKlahr2010}. Since we consider only the \textit{onset} of pebble accretion, the shear term, $\frac{3}{2}\Omega b$, can be ignored; then, $b_\mathrm{set}=\sqrt{4GM_p t_s/v_\mathrm{hw}}$. 

\begin{figure*}[t]
    \sidecaption
        \includegraphics[width=0.68\textwidth]{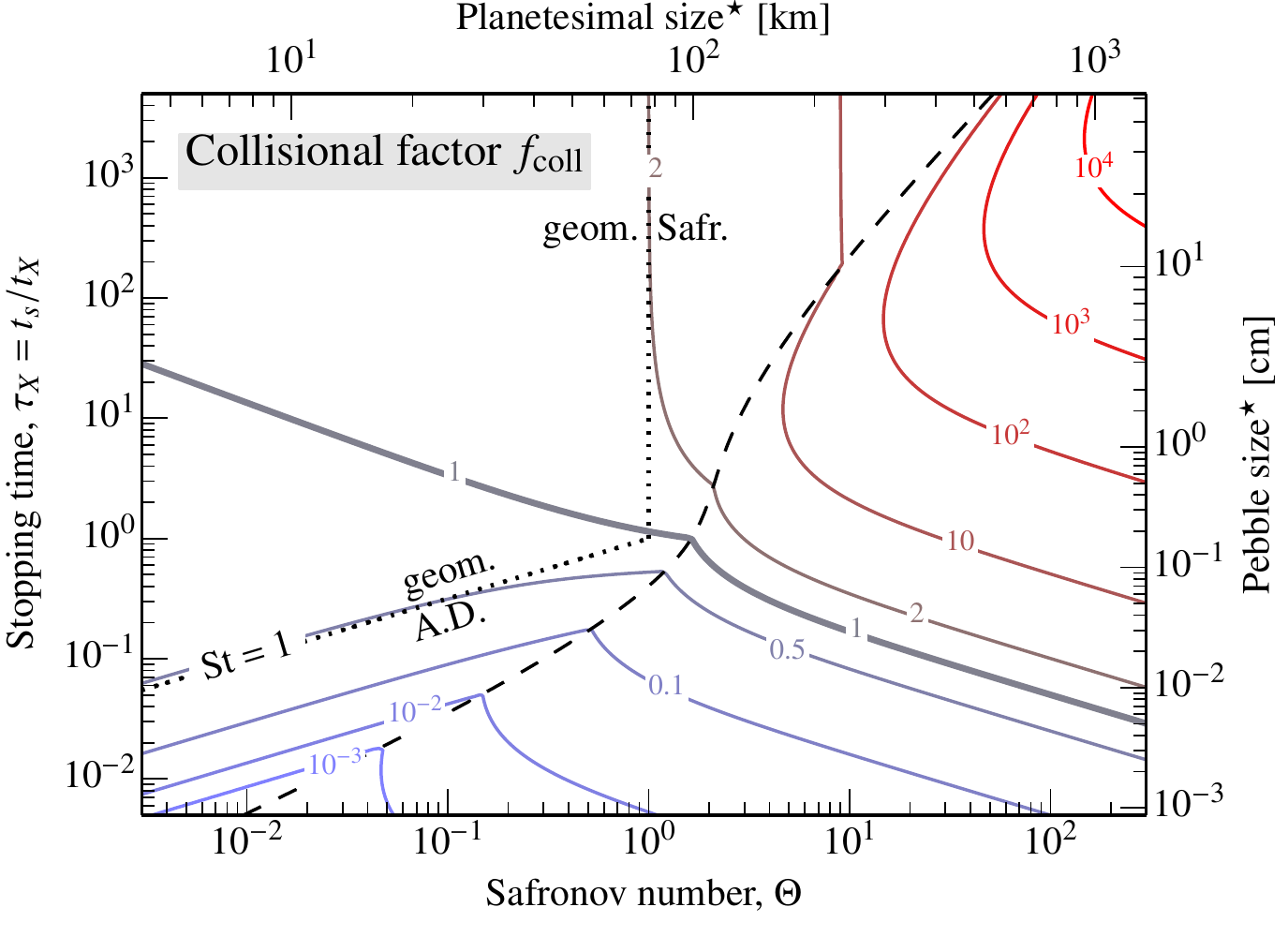}
    \caption{Contour plot of $f_\mathrm{coll}$ (\eq{fcoll-fit}) for potential flow as function of Safronov number (x-axis) and dimensionless stopping time $\tau_X = t_s/t_X$ (y-axis). The former is a proxy for the planetesimal size, whereas $\tau_X$ is a measure for the particle aerodynamical size. The dashed curve gives the transition between ballistic and settling interactions. The dotted lines $\Theta=1$ and $\mathrm{St}=1$ delineate the geometric regime (where $f_\mathrm{coll}\approx1)$ from the Safronov $\Theta>1$ and aerodynamical deflection (for $\mathrm{St}<1$ regimes).
    %Curves are labeled by the headwind Safronov number $\Theta=2GM_p/Rv_\mathrm{hw}^2$, which is a proxy for the mass of the planetesimal. 
    At small Stokes number aerodynamical deflection (A.D.) suppresses accretion, as these particles couple too well to the gas for ballistic interactions to be effective. 
    %On the other hand, the settling regime is curtailed for $\mathrm{St}>\mathrm{St}^\ast$ (open circles) as encounters become too fast for particles to reach their settling velocities. Thin, gray solid curves connect particles of the same stopping time $\tau_X = t_s\sqrt{8\pi G\rho_\bullet/3}$ (see \eq{tX}). 
    Particles of $\tau_X\lesssim1$ experience aerodynamic deflection \textit{before} entering settling regime, while $\tau_X\gtrsim1$ particles experience Safronov focusing before entering the pebble accretion regime.
    ($\star$: the conversion of $\tau_X, \Theta$ to physical parameters has been conducted for the default parameters at 1 AU.)
    }
%    \caption{Collision efficiency factor as function of Stokes number (x-axis) shown for the settling regime (solid lines) and the ballistic regime (dashed lines) for potential flow. Curves are labeled by the headwind Safronov number $\Theta=2GM_p/Rv_\mathrm{hw}^2$, which is a proxy for the mass of the planetesimal. At small Stokes number hydrodynamical effects suppress accretion, as these particles couple too well to the gas for ballistic interactions to be effective. On the other hand, the settling regime is curtailed for $\mathrm{St}>\mathrm{St}^\ast$ (open circles) as encounters become too fast for particles to reach their settling velocities. Thin, gray solid curves connect particles of the same stopping time $\tau_X = t_s\sqrt{8\pi G\rho_\bullet/3}$ (see \eq{tX}). With increasing planetesimal size they move from right to left in this plot. Particles of $\tau_X\lesssim1$ experience aerodynamic deflection \textit{before} entering settling regime, while $\tau_X\gtrsim1$ particles experience Safronov focusing before entering the pebble accretion regime.}
    \label{fig:analysis}
\end{figure*}
At large stopping times interactions are too fast for settling, because the interaction timescale \noeq{\sim}{b/\Delta v} becomes shorter than the stopping time scale and particles never reach terminal velocities \textit{during} the encounter. Consequently, \Eq{cubic} becomes invalid and pebble accretion inefficient for stopping times beyond a critical stopping time:
\begin{equation}
    t_\ast = \frac{4GM_p}{v_\mathrm{hw}^3},
    \label{eq:tast}
\end{equation}
implying that pebble accretion cross sections are largest for particles of stopping time $t_s\sim t_\ast$ \citep{LambrechtsJohansen2012}. 

For stopping times larger than $t_\ast$ particles experience the well-known (Safronov) gravitational focusing:  $b_\mathrm{Saf}=R\sqrt{1+\Theta}$, where $\Theta = (v_\mathrm{esc}/v_\mathrm{hw})^2$ is the Safronov number \changed{\citep{Safronov1972}} with $v_\mathrm{esc}=\sqrt{2GM_p/R}$ the surface escape velocity of the body. We refer to these encounters as `ballistic'
\footnote{Since the parameter space is enormous, gas drag-mediated accretion can be (sub-)categorized into several regimes. Originally, \citet{OrmelKlahr2010} identified  `settling', `hyperbolic' (=ballistic), and 3-body (more massive particles that experience little gas drag) encounters. \citet{LambrechtsJohansen2012} divided the settling regime into \changed{Hill accretion} (for massive bodies) and a \changed{Bondi} regime (used here). \mbox{\citet{GuillotEtal2014}} subdivided the ballistic regime into `geometric' (where $f_\mathrm{coll}\approx1$), `Safronov' ($f_\mathrm{coll}>1$), and `hydro', where $f_\mathrm{coll}\ll1$ due to aerodynamic deflection.}
because the encounter that grazes the object determines the impact parameter. This contrast the settling mode, where $b_\mathrm{set}$ is independent of the physical size $R$ -- the defining characteristic of pebble accretion. When encounters become too long gas drag acts to diminish the efficacy\? of ballistic interactions: projectiles follow gas streamlines.  Specifically, gas drag prevents small particles from colliding with the target when the stopping time is much less than $R/\Delta v_\mathrm{hw}$, or $\mathrm{St}\ll1$. Such particles never \textit{ballistically} collide with a planetesimal.

\begin{table}[t]
\centering
\begin{tabular}{lcccc}
\hline
\hline
 Flow & $f_\mathrm{coll,0}$ & $\mathrm{St}_\ast$ & $a$ & $b$   \\
 \hline
 \multicolumn{5}{c}{Settling interactions} \\
  Stokes    & $2\Theta \mathrm{St}$ & $2\Theta +4 +4/\Theta$ & $2.26$ & $0.61$ \\
 Potential  & $2\Theta \mathrm{St}$ & $2\Theta$ & $2.26$ & $0.61$ \\
 \hline
 \multicolumn{5}{c}{Ballistic interactions} \\
 Stokes & $1+\Theta$            & 1 & $3.24$ & $-0.86$ \\
 Potential & $1+\Theta$            & 1 & $0.78$    & $-0.89$ \\ 
 \hline 
\end{tabular}
\caption{Parameters for the collision factor for the settling and the ballistic regimes, see \eq{fcoll-fit}. Here, $f_\mathrm{coll,0}$ and $\mathrm{St}_\ast$ are given by physical considerations and $a$ and $b$ are fit parameters.}
\label{tab:fcolltab}
\end{table}
Thus, hydrodynamical effects suppress ballistic encounters below $\mathrm{St}\sim1$ whereas for larger Stokes numbers (stopping times) settling encounters vanish. We can describe both behaviors with a fitting function of the form:
\begin{equation}
    f_\mathrm{coll} = f_\mathrm{coll,0} \exp \left[ a\left(\frac{\mathrm{St}}{\mathrm{St}_\ast}\right)^b \right]
    \label{eq:fcoll-fit}
\end{equation}
where $f_\mathrm{coll,0}$ is the uncorrected collision efficiency and $a$ and $b$ are fit parameters that depend on the adopted flow pattern. The critical Stokes number $\mathrm{St}_\ast$ is fixed at 1 for ballistic encounters and $2\Theta$ (corresponding to $t_\ast$ in \eq{tast}) for settling interactions in potential flow. For Stokes flow we adopt a different prescription for $\mathrm{St}_\ast$. The parameters are listed in \Tb{fcolltab}.

These fits have been kept relatively simple and do not cover all situations.  For example, in the zero gravity limit ($\Theta=0$) $f_\mathrm{coll}$ does not become identically 0 below a critical Stokes number \citep{Slinn1976}. Generally, $f_\mathrm{coll}$ in the ballistic regime depends on the Reynolds number of the flow. For turbulent flows, furthermore, the suppression of $f_\mathrm{flow}$ at low St is much diminished (\citealt{HomannEtal2015}, see \se{flowdis}). In the settling regime, $f_\mathrm{coll}$ is consistent with earlier results \citep{OrmelKlahr2010,OrmelKobayashi2012} barring 3-body effects.\footnote{3-body effects become important for stopping times longer than the orbital period ($t_s \gtrsim \Omega^{-1}$) and for large planetesimals (protoplanets) for which $t_\ast \gtrsim \Omega^{-1}$. In these cases \eq{fcoll-fit} and \Tb{fcolltab} are not applicable.}

Contours of $f_\mathrm{coll}$ are given in \fg{analysis} for potential flow as function of Safronov number $\Theta$ and $\tau_X$, defined as
\begin{equation}
    \tau_X \equiv \Theta^{1/2}\mathrm{St} 
    = \frac{t_s}{t_X}
%    = \sqrt{\frac{8\pi G\rho_\bullet}{3}} t_s
    = \frac{t_s}{1337\ \mathrm{s}} \left( \frac{\rho_\bullet}{\mathrm{1\ g\ cm^{-3}}} \right)^{1/2},
    \label{eq:tX}
\end{equation}
where $t_X=\sqrt{3/8\pi G\rho_\bullet}$ depends only on the \textit{planetesimal} internal density. When $\rho_\bullet$ is constant, $\tau_X$ can therefore be identified solely with the pebble aerodynamical properties (in contrast to $\mathrm{St}$, which is a mix of planetesimal and pebble properties) allowing us to convert the dimensionless axes of \fg{analysis} to physical parameters. At high Stokes number (large $\tau_X$) and small $\Theta$ ballistic encounters operate in the geometric regime (top left). With decreasing $\tau_X$ (smaller particle sizes) interactions suffer  aerodynamical deflection as the Stokes number decreases. Collision efficiency factors then rapidly decrease, until the point where $f_\mathrm{coll}$ becomes determined by settling encounters (dashed curve). Nevertheless, for $\tau_X\lesssim1$ settling accretion rates are always modest even for $\Theta\gg1$ (bottom right corner). On the other hand, for $\tau_X \gtrsim 1$ particles the journey to pebble accretion involves the Safronov regime. This already ensures that $f_\mathrm{coll}$ increases above unity, but $f_\mathrm{coll}$ is especially boosted after the dashed curve has been crossed (top right).

%Ballistic encounters operate when $\mathrm{St}\gtrsim1$  and evaluate to focusing factors of ($1+\Theta$), independent of particle size. In contrast, settling encounters (solid curves) are effective at very small Stokes numbers, but they yield a much larger collision factor at higher St, because gas drag forces are less effective. Peak rates are obtained around $\mathrm{St}\approx\ \mathrm{St}_\ast=2\Theta$ (open circles), whereafter encounters become too fast for settling. Interactions then `return' to levels corresponding to the Safronov focusing of the ballistic regime.
%presents $f_\mathrm{coll}$ for potential flow as function of Stokes number for several Safronov numbers. 
%\footnote{For larger-mass planetesimals, that have $t_\ast \gtrsim \Omega^{-1}$, encounters will be determined by shear and no such decline takes place.}

These fits of \Tb{fcolltab} provide a good match to the data, see \fg{fcoll}. Note that $\mathrm{St}_\ast$ in the Stokes flow fit in the settling regime becomes $\gg$1 for small planetesimals ($\Theta \ll 1$). This extends the power-law solution of the settling regime and results in `flat' growth timescales (see \fg{Tgrowth}b). 
%For the Stokes flow we noted that settling encounters persisted to $\mathrm{St}>2\Theta$ at low $\Theta$. 
The physical reason behind this change is that for low impact parameters ($b\ll R$) encounter times are much longer than the $b_\mathrm{set,0}/v_\mathrm{hw}$ that is assumed in the derivation of $t_\ast$ (\eq{tast}), because the transverse velocity ($v_\theta$) approaches zero (see \fg{potstokes}). In fact, in Stokes flow the interaction time increases with decreasing cross section: $t_\mathrm{enc}\sim R/\delta v \sim R^2/b v_\mathrm{hw}$, where $\delta v$ is the transverse velocity at a distance $\delta r$ from the sphere and $\delta r \sim b$ (\textit{cf.}\  \citealt{JohansenEtal2015}). Continuing this order-of-magnitude analysis, one finds that the collision factor is still given by $b_\mathrm{set,0}^2$ but that the critical Stokes number (where $t_s=t_\mathrm{enc}$) is now $\mathrm{St}_\ast \sim \Theta^{-1}$. This explains the form for $\mathrm{St}_\ast$ adopted in \Tb{fcolltab}.

In the potential case a similar analysis results in $\mathrm{St}_\ast \simeq 1 +2\Theta$ but we found that the offset made little difference. More importantly, the fit expressions for potential flow fail at very small impact parameters (see \fg{fcoll}a). The reason is that centrifugal forces due to the flow curvature overwhelm gravitational forces, that is, particles are propelled away from the planetesimals, unless they collide head on. This results in another reduction of the collision efficiency compared to the fit in \fg{fcoll}a. However, we note that this is a rather academic result as for such small impact parameters these large velocities associated with the potential flow are anyway inappropriate (see \se{flowdis}).

\subsection{Critical stopping time for accretion}
From \fg{analysis} it is clear that the value of $\tau_X$ is a key factor in the growth of planetesimals.
%Although $\Theta$ is a proxy for the planetesimal's mass, St contains both planetesimal ($R$) and particle ($t_s$) properties. 
In \fg{analysis}, particles of the same stopping time move from left to right in \fg{analysis} as the planetesimal size increases.
%The significance of $\tau_X$ is the following. For any particle, a planetesimal can be found small enough such that interactions operate balistically at the geometric cross section of the planetesimal.
Starting from this point ($\Theta\ll1$ and $f_\mathrm{coll}=1$) and increasing the planetesimal size ($\Theta$) two growth modes can be identified: 
\begin{enumerate} 
    \item $\tau_X \ll 1$ particles become affected by  aerodynamic deflection (St $\lesssim1$) \textit{before} gravitational focusing kicks in ($\Theta>1$). With increasing planetesimal size (decreasing Stokes number), collision factors become significantly suppressed, $f_\mathrm{coll}\ll1$, stalling the growth of the planetesimal. At some point settling interactions will take over from ballistic encounters, whereafter $f_\mathrm{coll}$ increases with $\Theta$ in the settling regime. The settling regime accretion rates for $\tau_X\ll1$ particles are nevertheless modest:  $f_\mathrm{coll}\propto M_p$ (a.k.a. neutral growth). 
    \item $\tau_X \gg 1$ particles experience gravitational focusing ($\Theta>1$) \textit{first} and are never affected by aerodynamic deflection. Initially, $f_\mathrm{coll}$ increases due to (classical) Safronov gravitational focusing, $f_\mathrm{coll}\simeq\Theta$, but when $t_s\sim t_\ast$ at a Safronov number of $\Theta\sim\tau_X^{2/3}$, collision efficiencies exponentially increase as interactions transition to the (peak!) of the settling regime. Collisional focusing factors jump from $\sim$$\tau_X^{2/3}$ to $\sim$$\tau_X^{4/3}$ over a relatively short range in $\Theta$.
\footnote{The scalings follow by evaluating $f_\mathrm{coll}$ in the ballistic and settling regime for $\mathrm{St}=\mathrm{St}_\ast$, which occurs at $\Theta=(\tau_X/2)^{2/3}$. In reality due to the exponential tail, particles enter the settling regime at even lower $\Theta$, and the jump is even larger (see \eq{RPA}).}
For larger stopping times, the jump in $f_\mathrm{coll}$ associated with entering the pebble accretion regime is therefore more dramatic -- but occurs at a larger planetesimal mass.
\end{enumerate}

\section{Discussion}
\label{sec:Discussion}
\subsection{Viability of incremental growth of small planetesimals}
In our calculations of the growth timescale we assumed perfect sticking of pebbles with planetesimals. However,  dust collisional experiments often give a plethora of collisional outcomes, ranging from sticking, bouncing, erosion and catastrophic destruction \citep{Blum2008,GuettlerEtal2010}. This raises the question whether the reported growth timescales are realistic. Indeed \citet{Windmark2012} concludes that the (positive) mass transfer at small projectile sizes transitions to (negative) erosion if the projectile size becomes larger than \noeq{\sim}{0.1\ \mathrm{mm}}; and \citet{KrijtEtal2015} argues that erosion of large, fluffy planetesimals replenishes disks with a fresh reservoir in small grains.

\begin{figure}[t]
    \sidecaption
    \includegraphics[width=9cm]{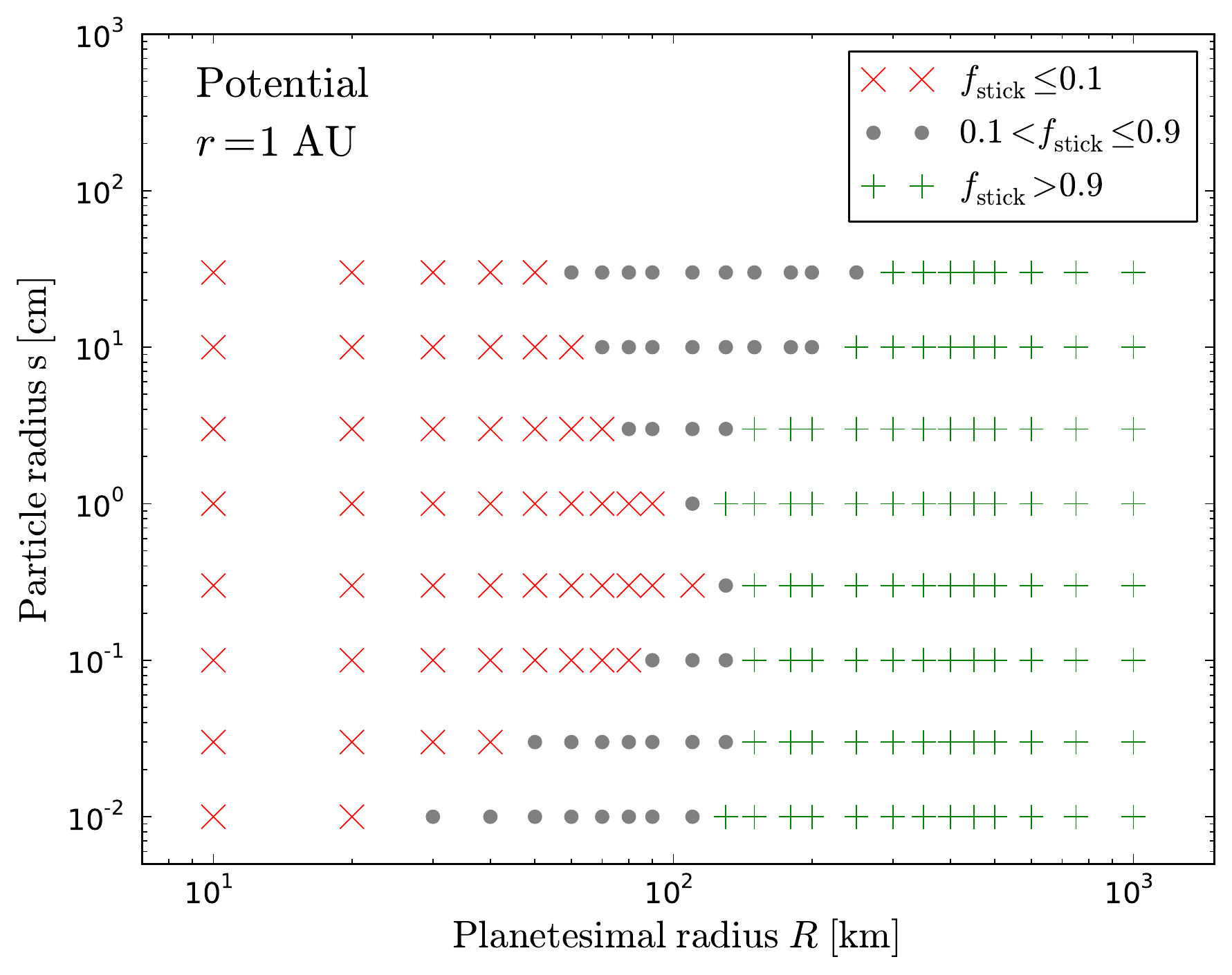}
    \caption{\changed{Scatter plot of planetesimal radius $R$ vs. particle radius $s$ with the potential flow at 1 AU from the star. The scatter plot shows the result of streamline integrations of the same parameter space as in \fg{fcoll}(a), assuming particles bounce off the surface. After 10 bounces particles are considered accreted. The red crosses represent a sticking fraction of  $f_\mathrm{stick} < 0.1$, the gray dots of $0.1 < f_\mathrm{stick} \leq 0.9$ and the green crosses $f_\mathrm{stick} > 0.9$.}}
    \label{fig:bouncefig}
\end{figure}

%, $b_\mathrm{coll,bounce}$, is less than $b_\mathrm{coll}$ determined to calculate the sticking efficiency factor:
%\begin{equation}
%    f_\mathrm{stick} =\left( \frac{b_\mathrm{coll,bounce} }{b_\mathrm{coll}}\right)^2,
%\end{equation} With $b_\mathrm{coll,bounce} = x_{2,B} - x_{1,B}$ and $x_{1,B}, x_{2,B}$ the first and last hit respectively after 10 bounces have been applied per particle streamline. 
\changed{To assess the dependence of the collisional physics on the growth rates, we considered a simulation where particles bounce off the surface of the planetesimal. We adopt the same parameters as in \fg{fcoll}(a) (1 AU, potential flow) and, following \citet{JohansenEtal2015}, adopted a coefficient of restitution of 50\% for a single bounce. Particles are considered accreted after they have bounced 10 times with the planetesimal. Clearly, the accretion cross section including bouncing is less than the previously determined collisional cross section. Let this ratio be called $f_\mathrm{stick}$. Hence, $f_\mathrm{stick}=0$ implies that all particles bounce off (to become re-entrained in the gas flow), whereas $f_\mathrm{stick}=1$ means that all colliding particles return to the planetesimal (because of its gravity). The results of these bouncing simulations are presented in \fg{bouncefig}, where symbols mark the value of $f_\mathrm{stick}$. The parameter space where the sticking fraction is low (red crosses: $f_\mathrm{stick} \leq 0.1$) by-and-large correspond to the geometrical regime. On the other hand, in the Safronov regime ($\Theta>1$), corresponding to a planetesimal size of 100 km, accretion becomes more effective, while in the settling regime $f_\mathrm{stick}=1$ is assured. In that case, gas drag and bouncing cause a sufficient amount of energy dissipation to gravitationally bind the pebbles. At small particle radii ($s = 0.01 \ \mathrm{cm}$ and $s = 0.03 \ \mathrm{cm}$), the sticking fraction again increases, because the few particles that accrete in this aerodynamical deflection regime do so by virtue of settling. All these findings are (qualitatively) consistent with \citet{JohansenEtal2015}, who considered a Stokes flow.}

%The geometrical regime persists longer for increasing particle size to 0.3 cm, where the geometrical regime is exceeded (see \fg{fcoll}(a)). This consequently increases the interval for which $f_\mathrm{stick} \leq 0.1$ for $s \leq 0.3$ cm.  This follows from the tangential impacts: the bounces are also approximately tangential and therefore do not influence $b_\mathrm{coll}$ to a high extend. Finally, The 2-body interactions ensure that despite the bouncing conditions, particles are still trapped and therefore $f_\mathrm{stick} > 0.9$ following up the transition interval of $0.1 < f_\mathrm{stick} \leq 0.9$. }

\changed{These results therefore show that planetesimals below 100 km cannot grow merely by gravitational effects.} However, the laboratory collision experiments that most resemble our case -- small projectiles that are being fired onto a solid wall -- \changed{present a more positive view}. \citet{Teiser2009} studied collisions between centimeter dust projectiles and a steel plate with velocities ranging from 30 to 47 $\mathrm{m \ s^{-1}}$. This experiment shows that the accretion efficiency is positive (30\%). 
%\citet{Windmark2012} concluded from collision experiments between submilimeter dust particles and kilometer sized objects that net growth is indeed a strong possibility. 
Furthermore, \citet{Meisner2013} studied the continuous impact of sub-millimeter (0.1--1 mm) dust particles onto larger objects. The results are consistent with the findings of \citet{Teiser2009} and show that planetesimals indeed gain net mass with an accretion efficiency of 30\% on average. 
%In addition, these experiments do not account for the self-gravity of the planetesimal, which, for a size of 100 km (where $t_\mathrm{growth}$ peaks), corresponds to an escape velocity of \noeq{\sim}{100\ \mathrm{m\ s^{-1}}}.
Therefore, collision physics is not necessarily an impediment for the sweepup growth scenario in the ballistic regime.

\changed{
But are these loosely-bound pebbles not simply blown off the surface ot the planetesimals as the latter moves through the nebular gas? On Earth, lift forces arise because of variations in flow velocities and pressure near the ground. On Earth, as well on Mars, lifting of small grains is the first step towards saltation -- a process that shapes desert landscapes and causes dust storms. Modelling these effects, \citet{ShaoLu2000} presented an empirical, but physically-motivated expression for the velocity required to initiate saltation (a.k.a.\ the fluid threshold; \citealt{Bagnold1936}):
\begin{equation}
    v^\ast \simeq 0.1 \sqrt{ \frac{2\rho_{\bullet,s} f_g s}{\rho_g} +\frac{\gamma}{2\rho_g s}},
    \label{eq:ShaoLu}
\end{equation}
where $v^\ast$ is the threshold velocity for lifting (which must be compared to the headwind velocity $v_\mathrm{hw}$), $f_g$ is the gravitational acceleration and $\gamma\sim0.1\ \mathrm{g\ s^{-2}}$ an empirical constant that represents the strength of the interparticle forces. The same forces operate on a rubble pile planetesimal moving through the nebular gas at velocities $v_\mathrm{hw}$. Therefore, when the magnitude of the lift force exceeds that of the gravity and interparticle forces, pebbles may be lifted from the planetesimal. \Eq{ShaoLu} suggests that small particles as well as large pebbles will stay bound because of sticking and high inertia, respectively. Nevertheless, for the relatively high densities in the inner disk $v^\ast$ may become less than the planetesimal headwind velocity for small planetesimals.\footnote{Great care must be taken to extrapolate \eq{ShaoLu}, which has been tailored for terrestrial conditions, towards the much rarefied nebular gas. \Eq{ShaoLu} must at least be modified in two points. First, it assumes a quadratic dependence on velocity of the gas drag law (as on Earth). Second, we can expect that small particles in the Epstein regime experience much reduced lift forces. }}

\subsection{Importance of the flow near the planetesimal}
\label{sec:flowdis}
In this work we have demonstrated that sweepup growth timescales increase for particles below a critical stopping time of $t_X\approx10^3$ s. In that case, $\mathrm{St}\ll1$ before Safronov focusing become important (\fg{analysis}). Under these conditions, we have seen that there is some difference in collision efficiency factors between the potential and Stokes flow solutions. Which of the two adopted solutions is more realistic?

Without access to hydrodynamical simulations, we can only be descriptive. However, as the planetesimal Reynolds number is $\gg$1, it may be argued that the potential flow solution -- as unphysical as it is -- is perhaps more applicable than the Stokes flow solution, which unrealistically alters the flow pattern out to several planetesimal radii (see our \fg{potstokes} and \citealt{Beard1974}). On the hand, at very small impact parameters ($f_\mathrm{coll}\ll1$) velocities close to the planetesimal surface must vanish, which renders Stokes solution more appropriate. Continuing this reasoning we may define a distance $\delta r$ from the surface of the planetesimal where the molecular diffusion timescale $(\delta r)^2/\nu$ equals the inertial transport time\? $v_\mathrm{hw}/R$. We then obtain $\delta r \sim R/\sqrt{\mathrm{Re}}$ (essentially the width of the boundary layer) as the point where the flow better resembles the potential solution than Stokes', with corresponding changes in the collisional efficiency of small particles.

%More importantly, for $\mathrm{Re}\gg10^2$\? the flow will become turbulent, which a qualitatively different outcome than the steady flows considered hitherto. In the zero-gravity limit the resulting collisional efficiency factors become additionally a function of particle Reynolds number $\mathrm{Re}$ (SLINN).

Another situation arises when the medium itself is turbulent. The gas in the protoplanetary disk, for one, is believed to be mildly turbulent due to hydrodynamic or magneto-hydrodynamical instabilities \citep{Balbus1991}. Recently, \citet{HomannEtal2015} numerically determined the collision efficiencies as function of Stokes number and turbulent intensity, finding a much shallower exponential decline (\textit{i.e.}, a fit parameter $a$ (see \Tb{fcolltab}) much closer to zero). This means, effectively, that the ballistic regime in \fg{analysis} becomes more important: the dashed line shift to the right `eating' into the parameter space where formerly settling encounters dominated. 

\changed{We reiterate that the above discussion only applies for the smallest particles ($t_s<t_X$); at $\sim$3 AU these are $\mu$m-size grains rather than mm/cm-size pebbles. Since these particles are accreted at $f_\mathrm{coll}\ll1$ efficiencies, meteorites can therefore only accrete trace amounts of them. While this is sufficient to explain the few presolar grains that are found in meteorites \citep{ClaytonNittler2004}, it is clear that accretion of micron-size grains will not significantly contribute to the total mass. Therefore, the bulk of the matrix material found in chondrites must have been brought in through larger, aerodynamically more weakly-coupled particles -- \textit{e.g.}\ as dust-rimmed chondrules or aggregates of dust-rimmed chondrules \citep{Cuzzi2004,OrmelEtal2008,JohansenEtal2015i}.}

Finally, while we have proposed a generally framework applicable to calculate accretion rates on planetesimals, it would be highly desirable to further refine our results by hydrodynamical simulations, which includes inertial particles that feel the gravity of the planetesimal.\footnote{Gravitational effects on the gas are of minor concern as long as the gas thermal motions are larger than the surface escape velocity.}

\subsection{Onset of pebble accretion}
In the discussion above we saw that, especially in the outer disk, peak growth timescales can easily become longer than the disk lifetime. This means that the fast pebble accretion regime is not accessible by incremental growth processes -- a situation exacerbated in the case when pebbles are drifting very quickly inwards and may be lost from the disk before the disk itself disappears. Therefore, a more attractive scenario for the growth of planets is to start from a planetesimal seed massive enough for interactions to fall in the pebble accretion regime. Such large planetesimals may be produced out of a collection of pebble-size particles by, \textit{e.g.}, streaming instabilities \citep{JohansenEtal2007,JohansenEtal2009,JohansenEtal2012}, being scattered from the inner disk, or be the collisional product of a population of small planetesimals\comm{refs}. However, this threshold for the onset of pebble accretion $R_\mathrm{PA}$ shifts to progressively larger sizes for increasing disk orbital radii and increasing particle stopping times (\Tb{fcolltab}).

We can obtain $R_\mathrm{PA}$ by equating $f_\mathrm{coll}$ in the Safronov and settling regime. However, this is a transcendental equation in $R$. To nevertheless obtain an (approximate) closed-form expression we use that the transition occurs very close to the point where $t_\ast \approx t_s$, yielding $R\approx v_\mathrm{hw} t_X (t_s/2t_X)^{1/3}$. Modifying this expression slightly, to be more in line with our numerical results, we find that the onset of pebble accretion occurs at a planetesimal radius:
\begin{align}
 R_\mathrm{PA} 
    &\approx 0.67 v_\mathrm{hw} t_X \left( \frac{t_s}{t_X} \right)^{0.28} 
    \quad \quad \qquad (t_s \gg t_X)
    \label{eq:RPA} 
    \\
    &\approx 520\ \mathrm{km}\
    \left( \frac{v_\mathrm{hw}}{\mathrm{50\ m\ s^{-1}}} \right)
    \left( \frac{\rho_\bullet}{\mathrm{g\ cm^{-3}}} \right)^{-0.36}
    \left( \frac{r}{\mathrm{AU}} \right)^{0.42} \tau_s^{0.28}
\end{align}
where the latter expression is valid for particles of $\tau_s = t_s\Omega <1$. \Eq{RPA} is
in good agreement with our numerical results.
%For an MMSN-disk and $\tau_s\simeq0.1$ pebbles this corresponds to Pluto-size objects ($R\approx10^3$ km) at 10 AU, consistent with our results. But at Kuiper-belt distances, $R_\mathrm{PA}$ approaches 2,000 km, explaining why Pluto stayed at its present size.

\section{Conclusions}
\label{sec:Conclusions}
In this paper we identify the onset of pebble accretion by finding the collision factor for $1$--$10^3 \ \mathrm{km}$ planetesimals at 1, 3 and 10 AU orbital distance for pebble sizes in the range of $0.01$--$30 \ \mathrm{cm}$. To investigate the influence of the flow pattern in the vicinity of the planetesimal, we have conducted numerical integrations for both the potential and Stokes flow solution. From our numerical integrations and analytical fits, we calculated a growth timescale on which the planetesimal mass $e$-folds by sweep-up of pebbles, investigating thereby the conditions under which small planetesimals can growth large enough for pebble accretion  to become important. We conclude the following from our study:  
\begin{enumerate}
      \item Gravitational interactions between pebbles and planetesimals ensure that the collisional efficiency factors never become zero, even for the smallest particles. %Although for these $f_\mathrm{coll} \ll 1$, gravitational settling always provide a floor to give $f_\mathrm{coll}>0$.
      \item Peak growth timescales for planetesimals sweepup of pebble-size particles occur at $\sim$100 km, corresponding roughly to the point where the headwind velocity equals the surface escape velocity of the planetesimal. \changed{This transition size therefore decreases when the disk headwind is lower, \textit{e.g.}\ for a colder disk.} For very small particles (very small stopping times) peak growth timescales are larger because of aerodynamical deflection.
%      \item In the potential flow solution, the peak growth timescale for growth by pebble sweepup occurs typically around $100 \ \mathrm{km}$. For the Stokes flow solution, the maximum growth timescale resembles\? a plateau and occurs at smaller planetesimal radius due to longer collisionally interaction timescales\? \comm{something like that}.
      \item For Stokes flow, where the gas velocities vanishes at the planetesimal surface, encounter times increase for smaller pebbles, increasing the likelihood to be captured by gravitational settling. In contrast, in potential flow pebbles are centrifugally ejected, strengthening the aerodynamic barrier.
      \item A critical particle stopping time of $t_X\approx 10^3\ \mathrm{s}$ (see \eq{tX}) distinguishes between the slow pebble accretion regime, where particles suffer from aerodynamic deflection, and the fast pebble accretion regime, where growth proceeds by ballistic encounters (Safronov focusing), before transitioning to the settling regime (pebble accretion).
      \item The onset of pebble accretion occurs when the planetesimal radius equals $R_\mathrm{PA}$ (\eq{RPA}). This is followed by an abrupt increase in the collision rate. Pebble accretion starts at larger planetesimal sizes, at increasing distance from the star.% since stopping times exceed encounter times when stopping times increase. 
      \item \changed{At $\sim$1 AU orbital distances, growth timescales are shorter than the disk lifetime of 10 $\mathrm{Myr}$ for any planetesimal size, provided pebble sizes are \noeq{>}{\mathrm{0.1\ cm}}. For small planetesimals (below 100 km) growth primarily depends on the collisional outcome (sticking, bouncing, or fragmentation) of the pebble-planetesimal collision.}
      \item \changed{In the outer disk collision timescales exceed the disk lifetime, implying that planetary seeds form from planetesimals large enough to gravitationally attract pebble-size particles: $R\gtrsim100$ km (Safronov focusing) or the faster pebble accretion whence $R>R_\mathrm{PA}$.}
\end{enumerate}

%Future work: stokes/potential realistic flow pattern (Homann) [Chris] 

\begin{acknowledgements}
The authors would like to thank Sebastiaan Krijt, Carsten Dominik, Tristan Guillot, Ralph Wijers, Lucia Klarmann for useful discussions and the referee, Anders Johansen, for an insightful report. C.W.O.\ is supported by the Netherlands Organization for Scientific Research (NWO).
\end{acknowledgements}

%-------------------------------------------------------------------
\bibliographystyle{aa}
\bibliography{aa}

\end{document}